\documentclass[fleqn,usenatbib]{rasti}

\usepackage{newtxtext,newtxmath}

\usepackage[T1]{fontenc}

\DeclareRobustCommand{\VAN}[3]{#2}
\let\VANthebibliography\thebibliography
\def\thebibliography{\DeclareRobustCommand{\VAN}[3]{##3}\VANthebibliography}

\usepackage[usenames,dvipsnames]{xcolor}
\usepackage{graphicx}	
\usepackage{amsmath}	
\usepackage{mathrsfs}
\usepackage{mathtools}
\usepackage{dsfont} 
\usepackage[normalem]{ulem}
\usepackage{siunitx}
\usepackage{xspace}
\usepackage{balance}
\usepackage{orcidlink}
\usepackage{booktabs}

\hypersetup{
    colorlinks = true,
    citecolor = {MidnightBlue},
    linkcolor = {BrickRed},
    urlcolor = {BrickRed}
}

\newcommand{\nside}{{\texttt{nside}}}
\newcommand{\lmax}{{\ensuremath{\ell_\text{max}}}\xspace}
\newcommand{\alm}{{\ensuremath{a_{\ell m}}}\xspace}
\newcommand{\D}{{\ensuremath{\mathfrak{D}}}}
\newcommand{\vecb}[1]{\boldsymbol{\mathbf{#1}}}
\newcommand{\case}[1]{\textbf{#1}}
\providecommand{\sorthelp}[1]{} 
\newcommand{\comment}[1]{{\color{black}#1}}

\title[Statistical estimation of full-sky radio maps using GCR]{Statistical estimation of full-sky radio maps from 21cm array visibility data\\ using Gaussian constrained realizations}

\author[K. A. Glasscock et al.]{
Katrine A. Glasscock$^{1}$\thanks{E-mail: katrine.glasscock@manchester.ac.uk (KAG)}\,\orcidlink{0000-0001-6894-0902},~
Philip Bull$^{1,2}$\,\orcidlink{0000-0001-5668-3101},~
Jacob Burba$^{1}$\,\orcidlink{0000-0002-8465-9341},~
Hugh Garsden$^{1}$\,\orcidlink{0009-0001-3949-9342},~
Michael J.\ Wilensky$^{1}$\,\orcidlink{0000-0001-7716-9312}~
\\
$^{1}$Jodrell Bank Centre for Astrophysics, University of Manchester, Manchester,  M13 9PL, UK\\
$^{2}$Department of Physics and Astronomy, University of Western Cape, Cape Town 7535, South Africa}

\date{Accepted 2024 October 3. Received 2024 September 27; in original form 2024 March 5}

\pubyear{2024}

\begin{document}
\label{firstpage}
\pagerange{\pageref{firstpage}--\pageref{lastpage}}
\maketitle

\begin{abstract}
An important application of next-generation wide-field radio interferometers is making high dynamic range maps of radio emission. Traditional deconvolution methods like CLEAN can give poor recovery of diffuse structure, prompting the development of wide-field alternatives like Direct Optimal Mapping and $m$-mode analysis.
In this paper, we propose an alternative Bayesian method to infer the coefficients of a full-sky spherical harmonic basis for a drift-scan telescope with potentially thousands of baselines \comment{, that can} precisely encode the uncertainties and correlations between the parameters used to build the recovered image. 
We use Gaussian constrained realizations (GCR) to efficiently draw samples of the spherical harmonic coefficients, despite the very large parameter space and extensive sky-regions of missing data. Each GCR solution provides a complete, statistically-consistent gap-free realization of a full-sky map conditioned on the available data, even when the interferometer's field of view is small. Many realizations can be generated and used for further analysis and robust propagation of statistical uncertainties.
In this paper, we present the mathematical formalism of the spherical harmonic GCR-method for radio interferometers. We focus on the recovery of diffuse emission as a use case, along with validation of the method against simulations with a known diffuse emission component.
\end{abstract}



\section{Introduction}
\label{sec:introduction}
Through measurements of the Cosmic Microwave Background (CMB), significant progress has been made in studying the very early Universe \citep{planck_2018}, but the subsequent epochs up to and including the Epoch of Reionisation still leaves much to be discovered.
A promising method to explore these intermediate epochs of cosmic history is the redshifted 21~cm line from neutral hydrogen. The signal can be used as a cosmological probe of structure formation, as it traces the distribution and evolution of the neutral hydrogen that fills the early Universe and forms the first stars and galaxies during Cosmic Dawn and also provides an inverse tracer of the reionisation of the intergalactic medium during the later Epoch of Reionisation (EoR) \citep{Pritchard_2012, Liu_2020}. Importantly, this tracer is redshift dependent, providing line-of-sight distance information that adds an extra dimension compared to the CMB, \citep{Furlanetto_2006,Morales_wyithe_2010,Liu_2020}.

Studies of the 21~cm signal are often divided into either measuring the spatially averaged \emph{global signal}, or measuring the statistical properties of the spatial fluctuations in the brightness temperature field. The 21~cm signal from the above-mentioned epochs is redshifted to the range $z\sim 27-6$, corresponding to frequencies of $50-\SI{200}{\mega\hertz}$. A number of experiments have been built with the purpose of searching for this signal, and many have already reported upper limits. A non-exhaustive list includes the Murchison Widefield Array; MWA \citep{bowman_MWA_2013,tingay_2013,wayth_2018}, the Donald C. Backer Precision Array to Probe the Epoch of Reionisation; PAPER \citep{Parsons_2012,ali_2015}, The LOw Frequency Array; LOFAR \citep{vanhaarlem2013,Patil_2017}, and the Hydrogen Epoch of Reionization Array; HERA \citep{DeBoer_2017,hera_collaboration_2022}.

The biggest technical challenge to date is the proper handling of foreground contaminants and how they are distorted by the receiving instrumentation. The dominant contaminant is typically the diffuse emission that is present at all parts of the sky, and is roughly a factor of around $10^5$ times brighter than the underlying 21~cm signal \citep{Liu_2020}. The main component of diffuse emission is the Galactic synchrotron emission, but there are also contributions from extragalactic free-free emission, bright Galactic radio point sources, and unresolved extragalactic point sources, \citep{santos_2005}. 
In addition there may be as-yet unknown diffuse components, e.g. associated with apparent excess background emission as claimed by ARCADE 2 \citep{Fixsen_2011, Singal_2010} and OVRO-LWA \citep{Dowell_2018_LWA}, or potential dark matter annihilation signals \citep{Evoli_2014_dm_annihilation,lopez-honores_2016_dm_annihilation}. Being able to robustly separate known but uncertain Galactic and extra-galactic components from novel sources of emission would greatly aid the physical interpretation of all of these effects, particularly as highly sensitive but complex next-generation arrays such as SKAO come online.

The issue of modelling diffuse emission is especially important for close-packed radio arrays. Sparse interferometers such as the \comment{Karl G. Jansky Very Large Array \citep[VLA;][]{Perley_2011_VLA} or SKA-MID \citep{Dewdney_2017_SKA-mid}} are essentially blind to large-scale emission, as the antenna configuration resolves out large angular scales. Compact arrays with smaller antennas are designed to make high dynamic range measurements on large angular scales however, which means the bright diffuse emission becomes an unavoidable issue. Imaging with interferometers is complicated, particularly for diffuse emission, as the effect of the interferometer response (its point spread function; PSF) must be deconvolved. For regular close-packed arrays, there is often highly incomplete coverage of the $(u,v)$-plane, since there are many baselines of the same length. This `redundancy' of the baselines results in artefacts like grating lobes in the PSF \citep{Dillon_2016}.

The typical lack of uniform $(u,v)$-coverage has given rise to the use of specialised deconvolution algorithms, where the so-called \emph{dirty image} is deconvolved to result in a more uniform image with imaging artifacts removed or suppressed. Many traditional pipelines make use of the CLEAN algorithm, which works by iteratively removing and replacing bright sources in the dirty image with their calculated point spread functions (PSFs) to get rid of the point source side lobes, \citep{Hogbom_1974_CLEAN,Cornwell_2009_comm_CLEAN}. CLEAN, however, does not correct for diffuse emission and additionally the statistics of the resulting image are not well known, particularly as CLEAN acts as a non-linear transformation of the data. 
Some improvements have been made to the CLEAN algorithm through Multi-Scale CLEAN \citep{Cornwell_2008_multiscale_clean} and WSCLEAN \citep{Offringa_2017_wsclean}, which can be used in either (or both) multi-scale and multi-frequency mode. Frequency dependence is introduced by dividing the full bandwidth into several output channels and then treating them individually. In multi-scale mode, further iterations are introduced using sub-loops and going through the scales one-by-one by convolving the dirty image with the corresponding space-kernel.
Even so, both methods still build upon the same CLEAN principle of using maxima to replace sources with an ideal PSF response. 

An alternative approach is Direct Optimal Mapping (DOM), which uses a maximum likelihood method to estimate the beam-weighted map of the sky using a pixel-basis, \citep{Dillon_dom_2015b,Xu_DOM_2022}. DOM is thus a one-shot process that estimates both the mean and the covariance in a lossless way, i.e. without losing information on model parameters. The model is built around a measurement matrix $\vecb{A}$ that maps a pixelated sky map to the visibilities. The choice of $\vecb{A}$ can be problematic however, and regularisation is needed in order to overcome degeneracies and ensure uniqueness of the map solutions that are obtained by `inverting' the measuring matrix. Furthermore, the noise covariance $\vecb{N}$ is also affected by the measurement matrix, requiring a choice for how to beam-weight the recovered map. 
While in principle the DOM formulation accounts for the whole sky, much smaller faceted maps are typically used to reduce the computational costs.

Like DOM, $m$-mode analysis is also typically implemented as a maximum likelihood estimator that will produce an estimate of the mean and covariance of the sky, although now in a spherical harmonic basis instead \citep{Shaw_2014, Shaw_2015, Eastwood_2018}. This method is developed specifically for drift-scan telescopes, taking advantage of simplifications that arise after applying a spherical transform to both the sky intensity and beam function, as well as performing a Fourier transform of the visibilities along the time direction, into $m$-mode space. As with DOM, proper care must be taken to ensure uniqueness of the solutions, so the inverse of the beam transfer function is replaced with the Moore-Penrose pseudo-inverse. Both DOM and $m$-mode analysis are maximum likelihood estimators, and therefore only provide moments of the underlying posterior distribution of the sky model parameters (although see Chapter~6 of \cite{PhD_Kriele_2022}, which recasts $m$-mode analysis as a Bayesian linear model).
\comment{ 
Other Bayesian approaches to radio interferometry analysis include
BIRO \citep{Lochner_2015_BIRO}, which is a Bayesian deconvolution algorithm that by-passes map-making by fitting models directly to the visibility data. BIRO, however, seems focused on modelling the flux densities of Gaussian sources on the sky, rather than being able to capture diffuse emission. Ref.~\citep{Roth_2023_resolve} also presents a Bayesian method, \texttt{resolve} that uses geometric variational inference to produce samples of the sky brightness distribution, given a model for the sky and antenna gains. Lastly, Bayesian methods have also been used to recover the 21-cm power spectrum, \citep[BayesEoR;][]{Sims_2016_BayesEOR, Burba_2023_BayesEOR}.
}

In this paper, we introduce a statistical method that, like $m$-mode analysis, is based on a spherical harmonic description of the sky (although we do not use an $m$-mode transform here). We use an explicitly Bayesian treatment of the map reconstruction problem, introducing a prior to the system which provides a regularisation, particularly in regions of missing data, and using the Gaussian constrained realization (GCR) method to make sampling of the large number of spherical harmonic coefficients tractable. Each sample vector drawn from the posterior is a valid realization of the whole sky, without any missing data regions, and we can treat it like we would treat `ideal' data without noise etc. Uncertainties can be propagated by passing multiple realizations through subsequent processing steps and then inspecting their distribution. By using an explicit forward model, we can also avoid ad-hoc beam weighting steps, which makes interpretation of the results more straightforward. Furthermore, this sampler can be incorporated in a larger `Gibbs sampling' framework that also samples other sky and instrumental parameters \comment{\citep[e.g.][]{Eriksen2008, Galloway_2023_commander_beyondplanck, Kennedy_2023, Amiri_2023}}.

Naturally, calculating multiple samples will require more computational power than solving once for the maximum likelihood (or for the maximum \emph{a posteriori}) solution. We propose some approaches to reduce the computational burden, such as specialising to the drift-scan telescope case and using Wigner \D-matrices to account for the sky rotating above the instrument. This removes the need to explicitly calculate the visibility response function (which maps the spherical harmonic coefficients to the visibilities) at multiple times. 

The paper is structured as follows: In Section~\ref{sec:bayesian_sampling} we build and present the visibility response model and cover the Bayesian methods of Wiener filtering and Gaussian constrained realizations. Both methods are related to Gibbs sampling, a future prospect of this work. Section~\ref{sec:simulations} describes the diffuse emission foreground model as well as the visibility simulation- and sampling parameters. In Section~\ref{sec:results} we present the performance for our reference (standard) case as well as a comparative analysis of various noise- and prior levels and ten other simulation scenarios including varying the field of view of the array. Finally, in Section~\ref{sec:conclusions} we conclude.

\section{Bayesian sampling of diffuse foregrounds}
\label{sec:bayesian_sampling}
In this section we develop the mathematical formalism to statistically sample spherical harmonic modes on the whole sky given radio interferometer visibility data. 

\subsection{Data model}
\label{subsec:data_model}
The data model in this work is based on a single diffuse component modelled as a sum of spherical harmonic modes at each frequency. Ultimately, this will only be one component of a more comprehensive sky model involving contributions from point sources, the EoR signal, etc. (which do not need to be modelled using a spherical harmonic basis). It would also be possible to define a particular functional form for the frequency dependence and interpret the spherical harmonic coefficients as amplitudes of this spectral template, e.g. by defining the specific intensity of the sky as
\begin{align}
    I(\nu, \theta, \phi) &= \sum_{\ell,m} a_{\ell m} Y_{\ell m}(\theta, \phi) f(\nu)
\end{align}
with frequency dependence $f(\nu)$. 
\comment{
In this analysis we only consider narrow bands of a few frequency channels in which we assume the sky emission to be effectively constant in frequency, i.e. we solve for a single set of spherical harmonics coefficients per narrow frequency bin of only two 1 MHz channels. It would then be possible to run the analysis independently for many such narrow bins and obtain posteriors for the SH coefficients in each of them, which could then be further studied, e.g. using a variety of spectral model fits, to estimate the frequency dependence of the sky signal. In this way, it would be possible to describe the temperature field with an arbitrary spatially varying frequency spectrum.
}
A common choice of frequency dependence is a power law with spectral index $\beta$.
These scenarios involve only simple modifications to the single component (and per-frequency) model that we consider here however, and so for the sake of simplicity we will not consider them further here.

We begin by writing out an expression for the complex visibilities observed by an interferometer. A baseline separated by vector $\vecb{b}$ given by antennas $i,j$ roughly probes the sky brightness temperature at Fourier mode $\vecb{u}=\vecb{b}/\lambda$, \citep{Liu_2020}. The visibility that the interferometer measures by that baseline $\vecb{b}$ is then given by,
\begin{align}
    V_{ij}(\nu,t) = \int\text{d}^2\Omega\, A_{i}(\nu,\vecb{\theta})A_{j}^{*}(\nu,\vecb{\theta})\, I(\nu,\vecb{\theta})\, e^\comment{-2\pi i \vecb{u}(\nu)\cdot \vecb{\theta}} + n_{ij},
\end{align}
where $A_{i}$ and $A_j^{*}$ are the E-field beams for each antenna, $I$ is the specific intensity of the sky, the exponential term describes the fringes where $\vecb{\theta}$ is the topocentric coordinates of the sources, and $n_{ij}$ is the baseline dependent noise.

The per-frequency spherical harmonic expansion of the sky model is given as,
\begin{align}
    I(\nu,\theta,\phi) = \sum_{\ell, m} a_{\ell m}(\nu)Y_{\ell m}(\theta,\phi),
\end{align}
where $\theta$ and $\phi$ are the declination and right ascension in equatorial coordinates respectively. 
For a typical drift scan array the motion of the sky is especially simple in an equatorial coordinate system, as it is a simple right ascension (or LST; \comment{Local Sidereal Time}) rotation.

There are many different conventions on which \alm modes to include, when working with spherical harmonics. For the sake of reducing computational resources, it makes sense to manipulate the \alm-vector to contain only the minimal required modes to solve the system. Firstly, it should be noted that even though the visibilities are complex entities, the actual sky needs to be a real field. The following symmetry relation for the \alm modes must be satisfied,
\begin{align}
    \alm = (-1)^m a^*_{l,-m},
\end{align}
or, split into real and imaginary parts,
\begin{align}
\label{eq:real-field_anti_symmetry_condition}
a_{\ell,+m}^{\rm re} = (-1)^m a_{\ell,-m}^{\rm re}\quad {\rm and}\quad
a_{\ell,+m}^{\rm im} = (-1)^{m+1} a_{\ell,-m}^{\rm im}\,.
\end{align}
This means that only $m\geq0$ modes are necessary and all negative $m$ modes are naturally excluded from the \alm-vector. 

Generally, the spherical harmonic coefficients are complex-valued. This can present complications when dealing with vectors of \alm-values numerically. We therefore split the \alm-values into their real- and imaginary parts. The \alm-vector now consists first of all the real- and then the imaginary modes. Moreover, since the $m=0$ imaginary parts will always be zero in order to satisfy the (anti-)symmetry condition of Eq.~\ref{eq:real-field_anti_symmetry_condition}, these spherical harmonic modes are removed while sampling and injected back in afterwards. For a given \lmax this leads to a total number of \alm modes of $N_\text{modes}=(\lmax+1)^2$. 

We next define the visibility response operator, $\delta V_{ij}^{\ell m}(\nu, t)$, which gives the projection from a spherical harmonic vector to a set of radio interferometer visibilities.
The visibility response is dependent on the antenna array configuration and location, which means the primary beam function is also absorbed into this operator. The full visibility will then be the visibility response function applied to the \alm modes summed over all $(\ell,m)$-values, 
\begin{align}
\label{eq:vis_response_as_sum}
    V_{ij} = \sum_{\ell m}\delta V_{ij}^{\ell m}(\nu, t)\,a_{\ell m}.
\end{align}
where the visibility response is defined as,
\begin{align}
\label{eq:visibility_response_full_definition}
    \delta V_{ij}^{\ell m}(\nu,t) =  \int\text{d}^2\Omega\, A_{i}(\nu,\vecb{\theta})A_{j}^{*}(\nu,\vecb{\theta})\, Y_{\ell m}(\alpha, \delta)\, e^\comment{-2\pi i \vecb{u}_{ij}(\nu)\cdot \vecb{\theta}}.
\end{align}
In this expression, $\vecb{\theta}$ are in topocentric coordinates (e.g. local altitude/azimuth), and $\alpha = \alpha(\vecb{\theta}, t)$ and $\delta = \delta(\vecb{\theta}, t)$ are the RA and Dec coordinates corresponding to a given topocentric pointing $\vecb{\theta}$ at local sidereal time $t$. This operator can be computed ahead of time for a given array configuration, which determines the available baseline vectors $\mathbf{u}$ and set of antenna E-field beams $A_i$.

\subsection{Wigner \D-matrix formalism}
\label{subsec:wigner_matrices}
In the above section, the visibility response is defined as a function of both time and frequency. 
Instead of simulating the visibility response operator for each time separately, we can choose a single reference time and apply a rotation to get the response at any other desired LST. This is made simpler by choosing the spherical harmonic basis to align with the rotation axis of the sky as seen by a drift-scanning telescope, i.e. by defining the spherical harmonic basis in equatorial coordinates. In this case, a simple azimuthal rotation around the celestial axis implements the mapping between RA and LST, while the declination stays constant.

We begin by defining the rotational matrix $\mathcal{R}_{\ell m\ell' m}$ that transforms the spherical harmonic modes at a reference sidereal time $t_\text{ref}$ to the updated sidereal time,
\begin{align}
\label{eq:alms_generic_rotational_matrix}
a_{\ell m}(t) &= \mathcal{R}_{\ell m\ell' m'}(t)\,a_{\ell' m'}(t_\text{ref}).
\end{align}
For spherical harmonics, this is given by the Wigner \D-matrix, which is a unitary matrix in an irreducible representation of the SO(3) group \citep{wu-ki_tung_group_theory}. The spherical harmonics transform as
\begin{align}
\label{eq:wigner_matrix_Y_lm}
    Y_{\ell}^{m}(\theta',\phi') &= \sum_{m'=-\ell}^{\ell}Y_{\ell}^{m'}(\theta,\phi)\,\D_{m'm}^{\ell}(\alpha,\beta,\gamma),
\end{align}
where $\D_{mm'}^{\ell}(\alpha,\beta,\gamma)$ is the \D-matrix given by the three Euler angles (that describe sequential rotations around three axes). Note, that for any rotation with \D-matrices, the spherical harmonic of degree $\ell$ and order $m$ transforms into a linear combination of spherical harmonics to the same degree $\ell$. The \D-matrix itself is given as 
\begin{align}
\label{eq:wigner_matrix_explicit}
    \D^{\ell}_{m' m}(\alpha,\beta,\gamma) &= e^{-i m' \alpha}\, d^{\ell}_{m' m}(\beta)\, e^{-i m \gamma},
\end{align}
where $d^{\ell}_{m' m}(\beta)$ are the corresponding reduced Wigner matrices. A table of the (small) $d$-matrices can be found in \cite{dong1994wigner}. Note the lack of mixing between $\ell$ modes.

For a zenith-pointing drift-scan telescope, the sky rotation in  a time interval $t-t_{\rm ref}$ can be mapped directly to an azimuthal rotation angle.
The full visibility given in Eq.~\ref{eq:vis_response_as_sum} can then be written as,
\begin{align}
\label{eq:visibility_with_wigner_sum}
    V_{ij}(\nu, t) = \sum_{\ell m}\delta V_{ij}^{\ell m}(\nu, t_\text{ref})\,\sum_{m'}\D^{\ell}_{m\,m'}(t)\,a_{\ell m'}(t_{\rm ref}),
\end{align}
where $\delta V_{ij}^{\ell m}(\nu, t_\text{ref})$ is the pre-computed visibility response at a reference time $t_\text{ref}$, $\D^{\ell}_{m\,m'}(t)$ is the appropriate \D-matrix for a given LST, and $a_{\ell m'}$ are the spherical harmonic modes of the sky at $t_\text{ref}$. If we let $p$ and $q$ label the frequency and time samples, $\nu_p$ and $t_q$, we can rewrite the above in index notation as
\begin{align}
\label{eq:visibility_index_notation}
    V_{ijpq} = {X}_{ij pq \ell m} \,\,a_{\ell m},
\end{align}
where repeated indices denote summation, and ${X}_{ij pq \ell m}$ is the combined operator for the two sums in Eq.~\ref{eq:visibility_with_wigner_sum}. The combined visibility response and \D-matrices now form the full $\vecb{X}$-operator, which is linear and equivalent to that of Eq.~\ref{eq:vis_response_as_sum}.

\subsection{Wiener filter -- the maximum a posteriori solution}
\label{subsec:wiener_filter}
Now, having established our visibility model based on the linear operator $\vecb{X}$ for the visibility response using a spherical harmonics basis --- we want to obtain the posterior distribution for the spherical harmonic modes given the data, noise, and priors, in order to generate samples of the diffuse emission sky. The steps outlined in this section and the next are steps of increasing complexity in a Bayesian hierarchy, ultimately preparing for implementing this model into a full Gibbs sampling scheme. Here, we go from the Wiener filter solution; a simple maximum a posteriori (MAP) solution, to drawing independent samples from the GCR equation. 

Thereby, as a first step in the Bayesian hierarchy we find the Wiener filter solution, which also acts as the central value for drawing the realizations. From \comment{Bayes'} theorem we can get the conditional distribution on the \alm modes,
\begin{align}
\label{eq:conditional_distribution_bayes}
    P(\vecb{a}\,|\,\vecb{d},\vecb{N},\vecb{a}_0,\vecb{S}) \propto \, &P(\vecb{d}|\,\vecb{a},\vecb{N})\,P(\vecb{a}|\,\vecb{a}_0,\vecb{S}),
\end{align}
when conditioning on the known covariances of the signal $\vecb{S}$ and noise $\vecb{N}$, the data-vector $\vecb{d}$, and the mean of the prior on the \alm modes $\vecb{a}_0$. For simplification of the notation we have dropped the $(\ell,m)$-indices and simply denote the \alm-vector as $\vecb{a}$ as well as dropping the baseline indices ($i,j$). It is assumed that generally the signal prior and covariance will be independent of the data and noise covariance. For simplicity we are keeping $\vecb{S}$ fixed, however, in a full Gibbs sampling scheme it would be possible to sample $\vecb{S}$ as well. 
Furthermore, we assume that the noise is Gaussian thus rewriting Eq.~\ref{eq:conditional_distribution_bayes} as:
\begin{align}
\label{eq:conditional_distribution_exponential}
     P(\vecb{a}\,|\,\vecb{d},\vecb{N},\vecb{a}_0,\vecb{S}) \propto e^{-(\vecb{d}-\vecb{X a})^\dag \vecb{N}^{-1}(\vecb{d}-\vecb{X a})} e^{-(\vecb{a}-\vecb{a}_0)^\dag \vecb{S}^{-1} (\vecb{a} -  \vecb{a}_0)}.
\end{align}
\comment{Note, that the data-vector $\vecb{d}$ refers to the complex-valued visibilities corresponding to the flattened real-valued \alm-modes.} For a Gaussian distribution the maximum of the posterior distribution is the same as the maximum of the log-posterior, hence we can find the Wiener filter solution by maximising the partial derivative of the exponent of Eq.~\ref{eq:conditional_distribution_exponential} with respect to the signal, $\vecb{a}$:
\begin{align}
     \left.\frac{\partial}{\partial \vecb{a}}\right|_{\vecb{a}=\vecb{\hat{a}}} \left(-(\vecb{d}-\vecb{X a})^\dag \vecb{N}^{-1}(\vecb{d}-\vecb{X a}) -(\vecb{a}-\vecb{a}_0)^\dag \vecb{S}^{-1} (\vecb{a} -  \vecb{a}_0) \right) = 0,
\end{align}
resulting in,
\begin{align}
     \comment{\vecb{\hat{a}}^\dag \vecb{X}^\dag \vecb{N}^{-1} \vecb{X}+ \vecb{\hat{a}}^\dag \vecb{S}^{-1} = \vecb{d}^\dag \vecb{N}^{-1}\vecb{X} + \vecb{a_0}^\dag \vecb{S}^{-1}.}
\end{align}
Now, taking advantage of the fact that covariance matrices are Hermitian, we can complex conjugate the entire expression and rearrange to obtain the Wiener filter solution \comment{$\vecb{a}_\textup{wf}$},
\begin{align}
\label{eq:wiener_filter_solution}
    \left[ \vecb{X}^\dag \vecb{N}^{-1} \vecb{X} + \vecb{S}^{-1} \right] \vecb{a}_\textup{wf} = \left( \vecb{X}^\dag \vecb{N}^{-1} \vecb{d} + \vecb{S}^{-1} \vecb{a}_0 \right) .
\end{align}
Returning to Eq.~\ref{eq:conditional_distribution_exponential}; multiplying two multivariate Gaussians will result in a new multivariate Gaussian.
By completing the square it can be shown that this can be written as proportional to a multivariate Gaussian with inverse covariance matrix given as
\begin{align}
\label{eq:inverse_covariance_wiener}
    \vecb{\Sigma}^{-1} = \vecb{X}^\dag \vecb{N}^{-1} \vecb{X} + \vecb{S}^{-1},
\end{align}
and where the Wiener filter solution acts as the mean,
\begin{align}
\label{eq:mean_wiener}
    \vecb{\hat{a}} = \vecb{\Sigma} \left( \vecb{X}^\dag \vecb{N}^{-1} \vecb{d} + \vecb{S}^{-1} \vecb{a}_0 \right).
\end{align}
Even though the Wiener filter is the maximum a posteriori solution, it is generally a biased estimator as described in more detail in \cite{Kennedy_2023}. Furthermore, the Wiener filter is a summary statistic but we are interested in generating actual samples that are complete and statistically-consistent realizations of the full-sky map. For this, the Wiener filter can instead act as the central value for drawing the samples.

\subsection{Gaussian constrained realizations}
\label{subsec:gaussian_constrained_realizations}
In order to draw samples from the full conditional distribution $P(\vecb{a}\,|\,\vecb{d},\vecb{N},\vecb{a}_0,\vecb{S})$, we can take advantage of the fact that we can describe our model as a multivariate Gaussian distribution. The Wiener filter solution described by its  mean and covariance given in eqs.~\ref{eq:inverse_covariance_wiener}~and~\ref{eq:mean_wiener} acts as a first step in the hierarchy to which we can add random normal realization terms of the signal $\vecb{\omega}_a$ and noise terms $\vecb{\omega}_d$ correctly scaled by their respective covariances to draw the constrained realizations $\vecb{a}_\textup{cr}$, \citep{Eriksen2008}. This leads to the GCR equation,
\begin{align}
    \label{eq:GCR_simple_version}
    \left[ \vecb{X}^\dag \vecb{N}^{-1} \vecb{X} + \vecb{S}^{-1} \right] \vecb{a}_\textup{cr} =  \left( \vecb{X}^\dag \vecb{N}^{-1} \vecb{d} + \vecb{S}^{-1} \vecb{a}_0 + \vecb{S}^{-\frac{1}{2}} \boldsymbol{\omega}_a + \vecb{X}^\dag \vecb{N}^{-\frac{1}{2}} \vecb{\omega}_d \right).
\end{align}
In turn Eq.~\ref{eq:GCR_simple_version} reduces back to the mean $\langle \vecb{a}_\textup{cr} \rangle = \vecb{\hat{a}}$, since the unit variance Gaussian vectors $\vecb{\omega}_a$ and $\vecb{\omega}_d$ have zero mean. By solving Eq.~\ref{eq:GCR_simple_version} repeatedly we can draw independent samples of the \alm modes on the sky consistent with the given data vector $\vecb{d}$, chosen signal prior $\vecb{a}_0$, the visibility response operator $\vecb{X}$, and the covariances $\vecb{S}$ and $\vecb{N}$. For the sake of keeping computational costs low, we have kept to \comment{200} samples per set of parameters for this paper. 

In the case studied in this paper, there is only missing data outside of the FOV defined by the primary beam and the horizon; both of which is dealt with through the visibility response function. In case of actual gaps in the data, for instance due to RFI flagging, one can define a set of weights $\vecb{w}$ to redefine the inverse noise covariance,
\begin{align}
\label{eq:noise_covariance_weighted}
    \vecb{N}_{\vecb{w}}^{-1} = \vecb{ww}^T \circ \vecb{N}^{-1},
\end{align}
where $\circ$ denotes element-wise multiplication. Using this weighted noise covariance in Eq.~\ref{eq:GCR_simple_version} means we automatically zero the contribution from the data inside the mask. The signal covariance, however, does not necessarily go to zero inside the masked regions and thereby \emph{takes over} the signal estimation in the absence of information from the likelihood function. Note that this is not equivalent to drawing samples from the prior distribution and simply filling-in the masked regions. The samples generated within the flagged regions will be informed by both the prior {\it and} data from the un-flagged regions. 

In the future this process will be done within the context of Gibbs sampling \citep{geman1984}, which is a Bayesian method to sample directly from the joint posterior of more complicated high-dimensional distributions. In this instance, we focus only on the distribution for the spherical harmonic coefficients, and use the conjugate gradient solver \comment{from the sparse linear algebra module from \texttt{scipy} \citep{2020SciPy-NMeth}} to obtain $\vecb{a}_\textup{cr}$ from the GCR equation. Most solvers do not handle complex quantities well so to overcome this, we saw it necessary to `realify' the full system as described in App.~\ref{app:realification}.

When implemented into a Gibbs sampling scheme, it will enable us to also sample the signal covariance $\vecb{S}$. Gibbs sampling is an iterative method, that samples from each conditional distribution in turn and thereby effectively samples from the joint posterior. For each iteration it updates the conditional variables with the samples obtained from the previous step,
\begin{subequations}
\begin{align}
    \label{eq:gibbs_sampling_scheme_top}
    \vecb{a}_{i+1} &\leftarrow P\!\left(\vecb{a}_i \,|\, \vecb{d}, \vecb{N}, \vecb{a_0},\vecb{S}_i\right), \\
    \vecb{S}_{i+1} &\leftarrow P\!\left(\vecb{S}_i \,|\, \vecb{a}_{i+1}\right),  
    \label{eq:gibbs_sampling_scheme_bottom_Signal_cov_conditional}
\end{align}
\end{subequations}
where $\leftarrow$ represents generating samples through evaluating the conditional distribution and $i$ indexes the Gibbs iteration number. First, the $\alm$ modes are sampled and it is noticed that the conditional distribution takes the same form as in Eq.~\ref{eq:conditional_distribution_bayes}. We can therefore use the same arguments as before; we are still dealing with a multivariate Gaussian distribution and therefore end up at the GCR equation of Eq.~\ref{eq:GCR_simple_version}. The sample obtained from Eq.~\ref{eq:gibbs_sampling_scheme_top} already contains the conditioning on the noise covariance $\vecb{N}$, data $\vecb{d}$, and prior mean $\vecb{a_0}$, so there is no need to condition again on these parameters when sampling $\vecb{S}$ in Eq.~\ref{eq:gibbs_sampling_scheme_bottom_Signal_cov_conditional}. We leave a full exploration of sampling the signal covariance (which encodes the spherical harmonic angular power spectra) to later work.

\section{Simulations and sky model}
\label{sec:simulations}
The structure of the simulated visibility response function has already been laid out in Sec.~\ref{subsec:data_model} where it is also made clear that the simulation is dependent on which LSTs are chosen, \lmax, frequency $\nu$, baseline orientation and length $b$, and dish size $\theta_D$. This section will describe how the simulations were obtained as well as the standard parameters used for the reference results, which from here on will be referred to as the \emph{standard} case. More realizations have been drawn using a variation of both the simulation/observation parameters as well as samples from using a variation of noise and prior levels. The details of these parameter variations are explained in Sections~\ref{subsec:noise_and_prior_levels}--\ref{subsec:beam_3_m_mwa_like}. 

\subsection{Precomputing the visibility response operator}
\begin{figure*}
 	\includegraphics[width=\textwidth]{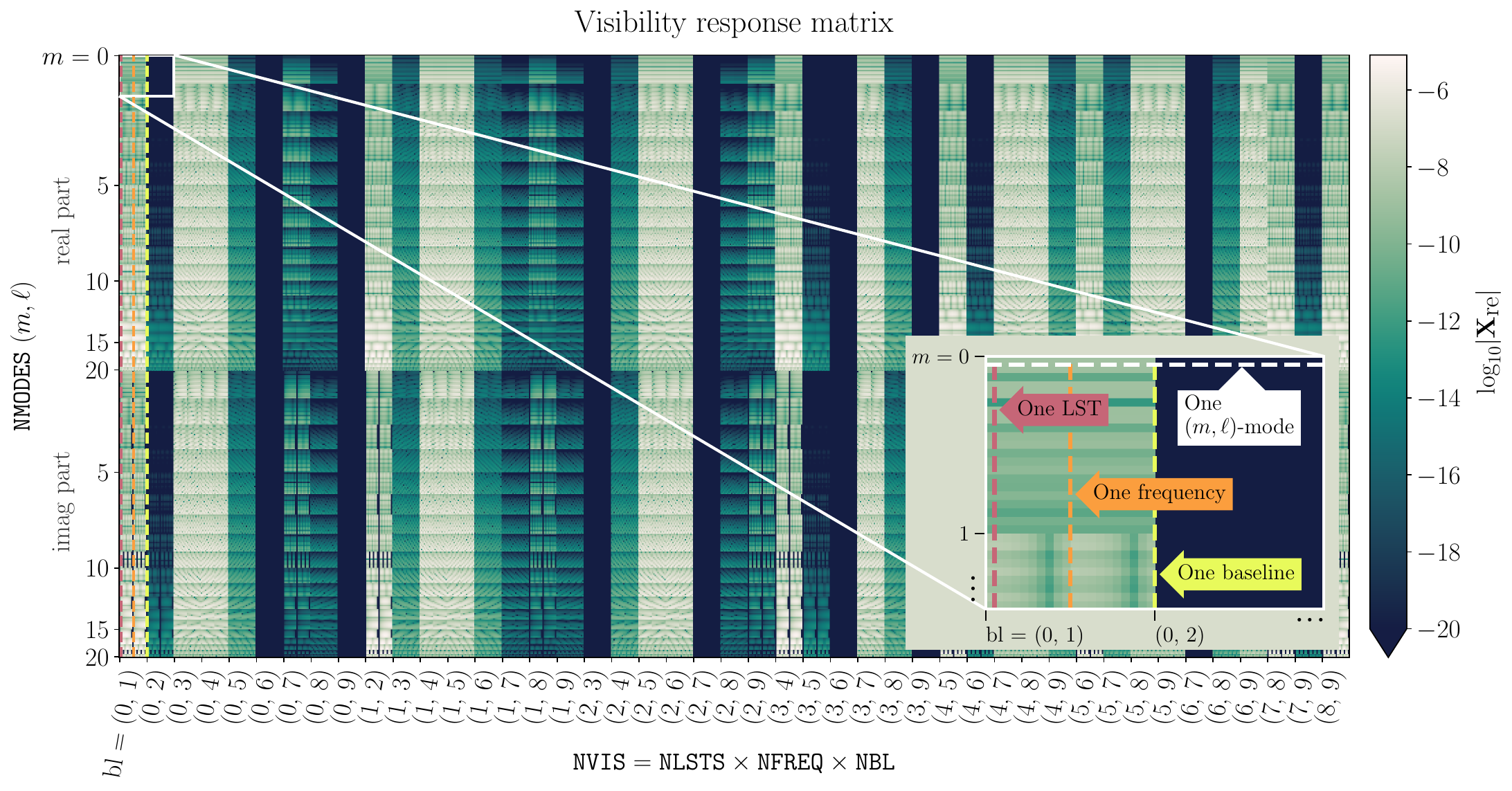}
    \caption{The absolute values of the visibility response operator $X_\text{re}$ on a log-scale and with dimensions (\texttt{Nmodes}) $\times$ (\texttt{NLSTs} $\times$ \texttt{Nfreq} $\times$ \texttt{Nbl}). Here it is shown for $\ell_\text{max}=20$, LST $=$ 0--8~\si{\hour},  $\nu =$ 100--101~\si{\mega\hertz} and 10 close packed antennas that form 45 baselines as shown in Fig.~\ref{fig:array_antenna_config}. The darker is is, the less that frequency/LST/baseline contributes to measuring that $(m,\ell)$-value. Note that this is the \emph{real-part only} of $X$ (the imaginary part showing similar structure) but it covers both the real- and imaginary parts of the \alm modes.}
    \label{fig:vis_response_operator}
\end{figure*}

To simulate the visibility response function we used the \texttt{hydra}\footnote{\url{https://github.com/HydraRadio/}} code, which builds upon \texttt{matvis} \citep{matvis_viscpu_2023} to simulate the visibility per \alm-mode as well as per baseline, frequency, and LST. The absolute values of the real-part of the visibility response operator, $\vecb{X}_{\rm re}$, is shown in Fig.~\ref{fig:vis_response_operator} also showing how the operator's dimensions are packed. The visibility response is only simulated for Stokes-I. We have assumed the most general case where we want to solve for a different set of $\alm$ modes for each frequency channel. The simulations cover the narrow bandwidth of \num{100}--\SI{102}{\mega\hertz} in just two channels, in order to keep the problem small for the time being. 
The LST range spans \num{0}--\SI{8}{\hour} in 10 steps. The primary beam model is calculated using \texttt{pyuvsim}\footnote{\url{https://github.com/RadioAstronomySoftwareGroup/pyuvsim}} which provides an analytic Gaussian beam model assumed to be identical for each antenna, with the width $\theta \simeq \lambda/b$ where $b$ is length of the baseline $\vecb{b}$. As standard the distance between each antenna is \SI{14.6}{\meter} (HERA-like) and the dishes themselves are modelled as hyperbolic dishes of diameter $\theta_\text{D}=\SI{14}{\meter}$ with no side-lobes. The model is simple, but has been specifically chosen so that we can clearly understand and validate the results. The impact of different beams is an interesting study in its own right, which we defer to future work.

\begin{figure}
	\includegraphics[width=\columnwidth]{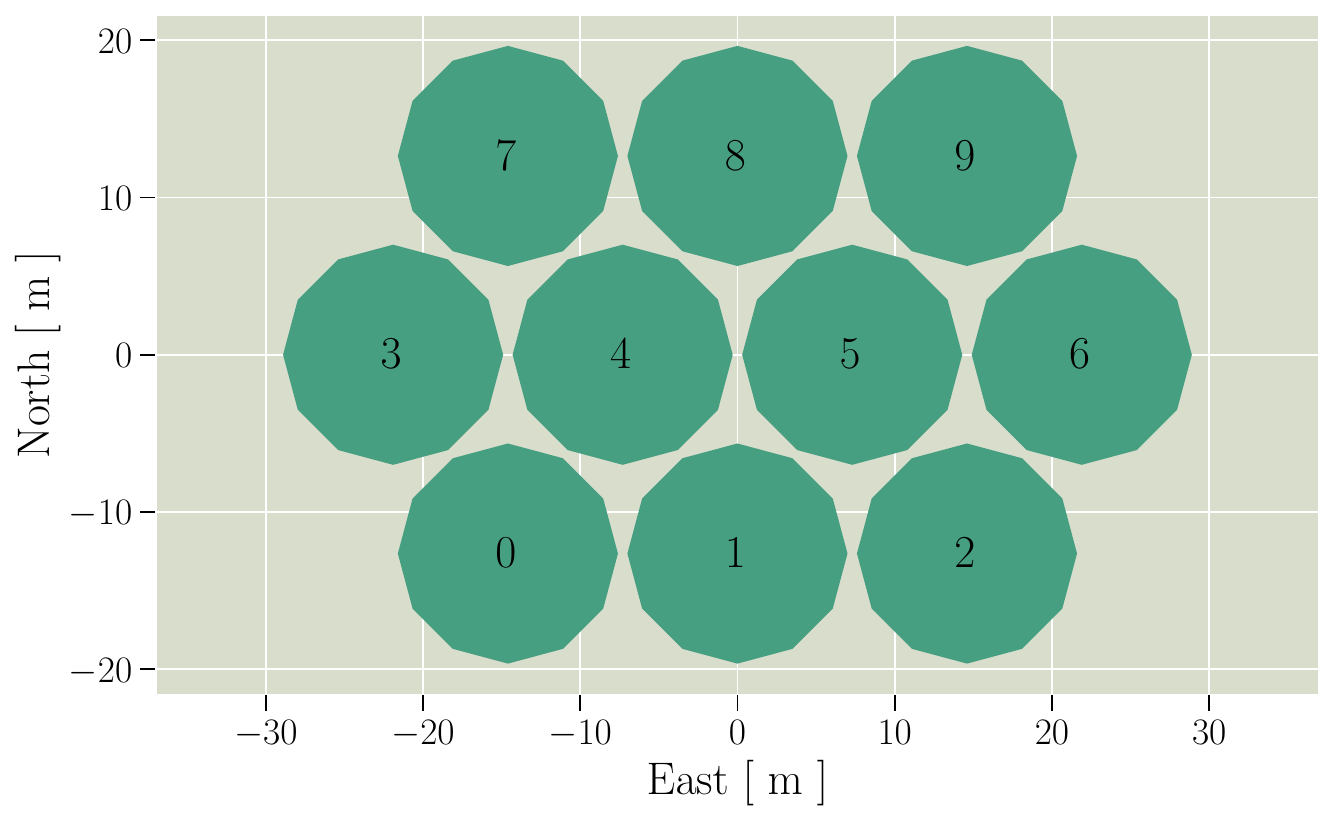}
    \caption{The standard array layout used in the majority of the simulations, consisting of 10 closed packed antennas with a diameter of \SI{14}{\meter} and a separation of \SI{14.6}{\meter}.}
    \label{fig:array_antenna_config}
\end{figure}

The standard array layout used in the simulations is a small closed-packed redundant array to represent a subsection of the full HERA array. A redundant layout benefits from sampling the same modes many times, thus increasing sensitivity especially to the relatively large angular scales relevant for EoR experiments. In this configuration we consider only 10 receivers as shown in Fig.~\ref{fig:array_antenna_config}. The subset is sufficiently small to both keep computational costs low and with its resulting 45 baselines it still offers multiple redundant baseline groups and variation in both direction and baseline length. As with HERA the array used in the simulations is a drift-scan instrument pointing towards zenith placed at latitude $\sim\SI{-31}{\degree}$. From Fig.~\ref{fig:vis_response_operator} it is already clear that some baselines contribute more than others. For instance the baseline of antennas $(3,6)$ is not very responsive, which is the long E-W $\SI{43.8}{\meter}$ baseline (Fig.~\ref{fig:array_antenna_config}) whereas all the short $\SI{14.6}{\meter}$ baselines contribute significantly more.

\subsection{Diffuse emission sky model}
\label{subsec:diffuse_emission_sky_model_pygsm}
As we aim to sample the diffuse emission foregrounds specifically, no other foregrounds are included in the foreground model. For the purely diffuse emission foreground sky model we use the Global Sky Model \citep[GSM2016;][]{global_sky_model_2016} as implemented by \texttt{pyGDSM} \citep{pyGDSM_ascl_ref_price_2016}. GSM2016 is based on principal component analysis of a large set of multi-frequency datasets (spanning \SI{10}{\mega\hertz} to \SI{5}{\tera\hertz}) as well as performing a blind separation into physical components revealing five components identified as synchrotron emission, free-free emission, cold dust, warm dust, and the CMB anisotropy. The sky model is provided as frequency dependent \texttt{HEALpix}\comment{\footnote{\url{http://healpix.sourceforge.net}}} maps where we use only the GSM2016 map at reference frequency \SI{100}{\mega\hertz}. The foreground map is then passed to \texttt{HEALpy}, \comment{\citep{Gorski2005HEALpix,Zonca2019HEALpy}}, to get the true \alm modes, $\vecb{a}_\text{true}$. As standard, the resolution is set to \nside~$=128$, which corresponds to a HEALPix angular pixel size of $\SI{0.46}{\degree}$. The maximum mode of the operator is set to $\lmax=20$, however, which corresponds to $\sim \SI{9}{\degree}$. 
The true sky is used as input for the full data model, which is obtained by applying the visibility response to the true \alm modes.

\subsection{Data and noise model}
\label{subsec:data_and_noise_model}
The data input for the analysis in this work is based on the visibility response simulations. In the future, this will be done with real visibility data instead, but for now simulations with a known diffuse sky component serve as a means of validation. Here, we use the visibility response operator $\vecb{X}$ to map the true-sky spherical harmonic coefficient vector $\vecb{a}_{\rm true}$ into visibilities and add a noise vector $\vecb{n}$ to the simulated visibilities,
\begin{align}
\label{eq:data_input}
    \vecb{d} = \vecb{X} \vecb{a}_\text{true} + \vecb{n}. 
\end{align}

The noise on the data model is added as uncorrelated \comment{complex} Gaussian random noise, \comment{$\vecb{n} = \vecb{n}_\text{re} + \vecb{n}_\text{im}$, where }
\begin{align}
\label{eq:uncorrelated_gaussian_random_noise}
\comment{\vecb{n}_\text{re}, \vecb{n}_\text{im}} \sim \mathcal{N}(0,\vecb{N}),
\end{align}
given by the noise covariance $\vecb{N}$ with components $N_{ij}$ modelled by the simulated auto-correlation visibilities $V_{ii}$ and $V_{jj}$ as given by the radiometer equation,
\begin{align}
\label{eq:radiometer_equation}
    N_{ij} = \sigma_{ij}^2 = \frac{V_{ii}V_{jj}}{\Delta t \Delta \nu}.
\end{align}
Since the foregrounds are spectrally smooth, there is no need for very high spectral resolution and we choose $\Delta\nu = \SI{1}{\mega\hertz}$. The time resolution is set to $\Delta t = \SI{60}{\second}$, both to avoid issues with sky rotation (smearing) and to still ensure good signal-to-noise ratio.

Lastly, we have applied a $10\%$ prior on the $\alm$ values. To implement this, we have defined a prior covariance $\vecb{S}$ that is diagonal, with values ${S}_{nm}=(0.1\,a_{\text{true},n})^2 \delta_{nm}$. The prior mean, $\vecb{a}_0$, is not set equal to $\vecb{a}_\text{true}$ however, as this would effectively be inputting information about the correct answer into the inference ahead of time, which is not realistic. Instead, we draw values for $\vecb{a}_0$ from a Gaussian distribution centered on the true \alm values, with the same prior covariance, i.e. $\vecb{a}_0\sim \mathcal{N}(\vecb{a}_\text{true}, \vecb{S})$. 
This ensures that the prior mean that we input in the GCR equation (Eq.~\ref{eq:GCR_simple_version}) is set consistently with the chosen prior model (\texttt{GSM2016}), while deviating from the true values of the parameters (as would be the case in a real analysis). 

\begin{figure*} 
	\includegraphics[width=\textwidth]{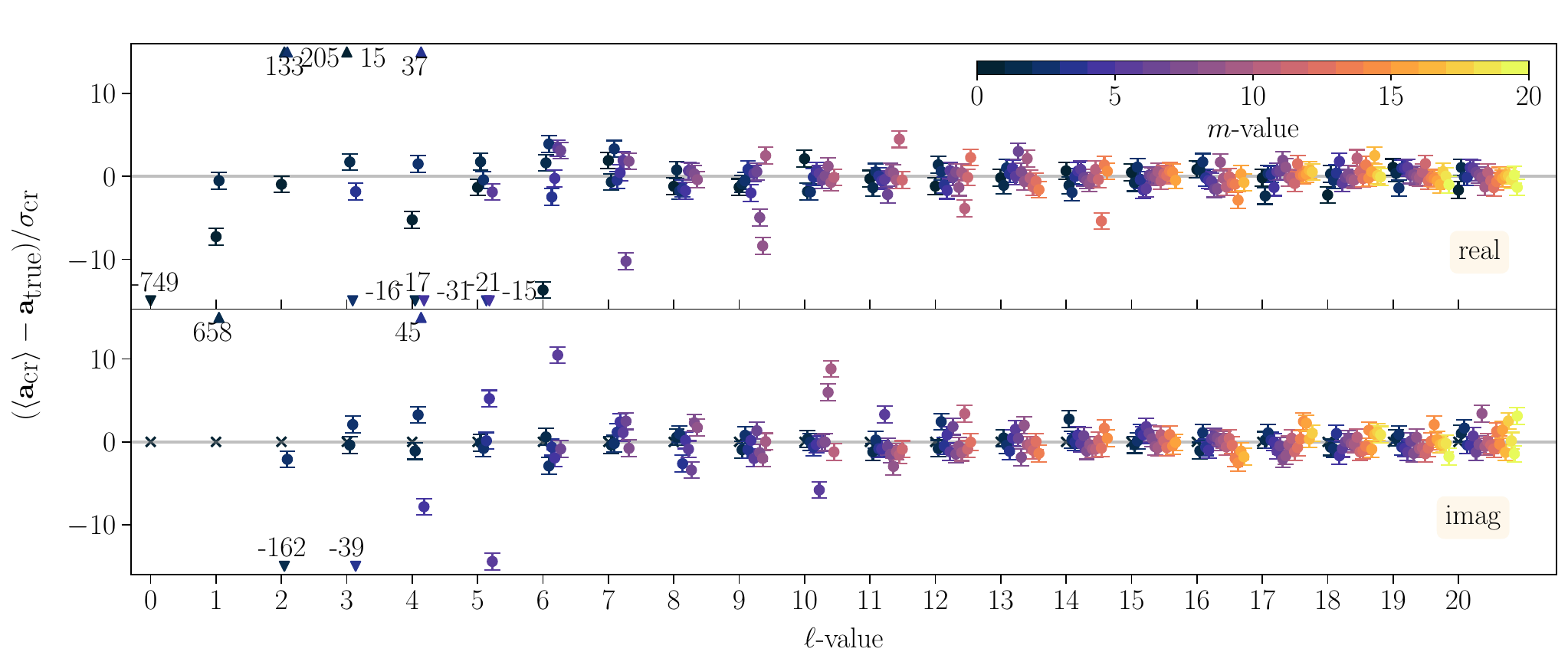}
    \caption{The real- and imaginary parts of the difference of the mean of \comment{200} samples of \alm modes from the GCR solver and the true sky using the \emph{standard} configuration and parameters normalised by the individual standard deviations. Any \emph{outliers} from the central region is marked with $\blacktriangle$ and it is notable that these occur more frequently in the low-$\ell$ region. As described in Sec.~\ref{subsec:data_model} the $m=0$ imaginary modes ($\times$) should always be zero and the GCR solver therefore does not solve for this.
    }
    \label{fig:alms_standard}
\end{figure*}

\begin{figure*}
\centering
	\includegraphics[width=0.95\columnwidth]{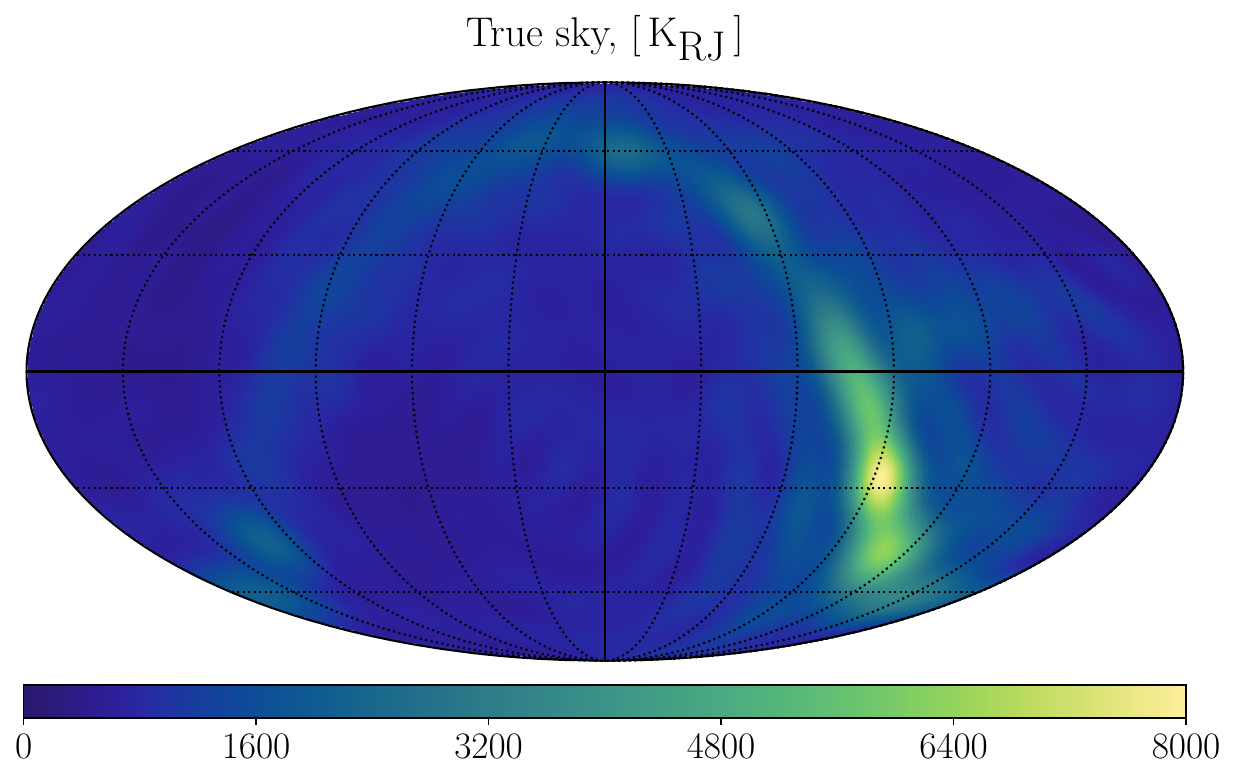}\hspace{0.06\columnwidth}
 	\includegraphics[width=0.95\columnwidth]{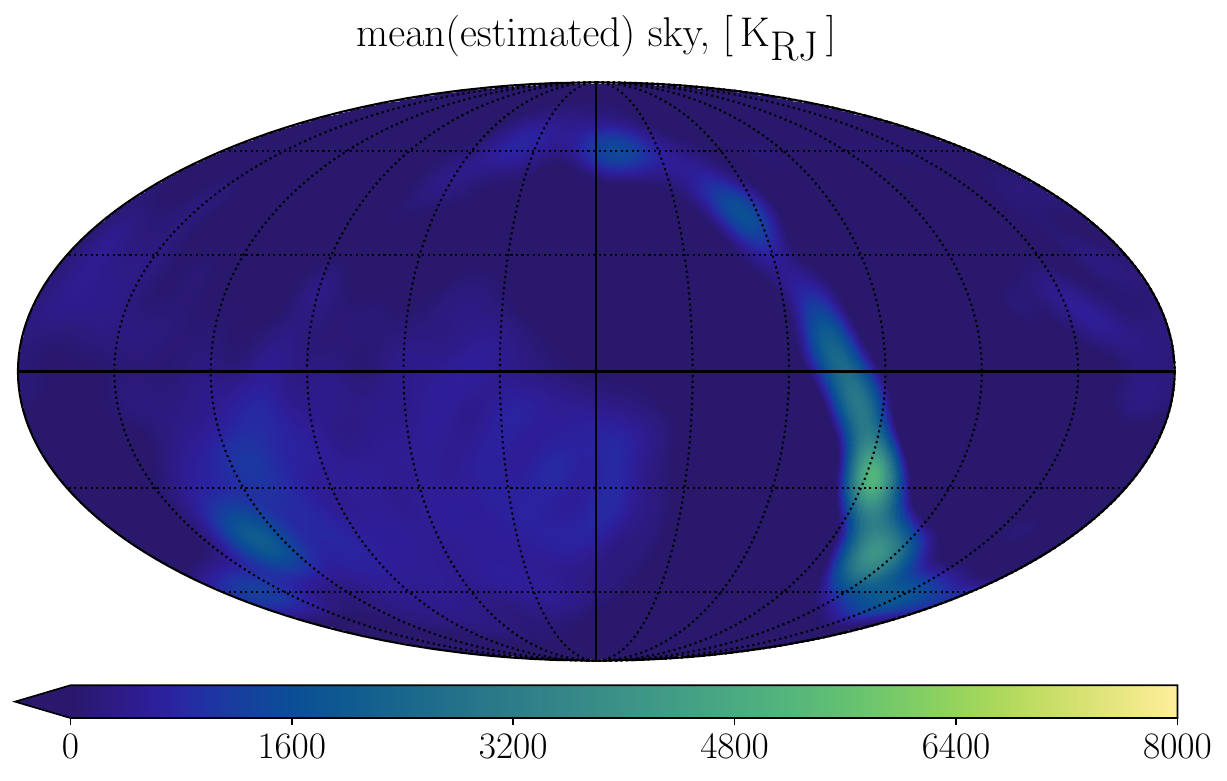}
    \includegraphics[width=0.95\columnwidth]{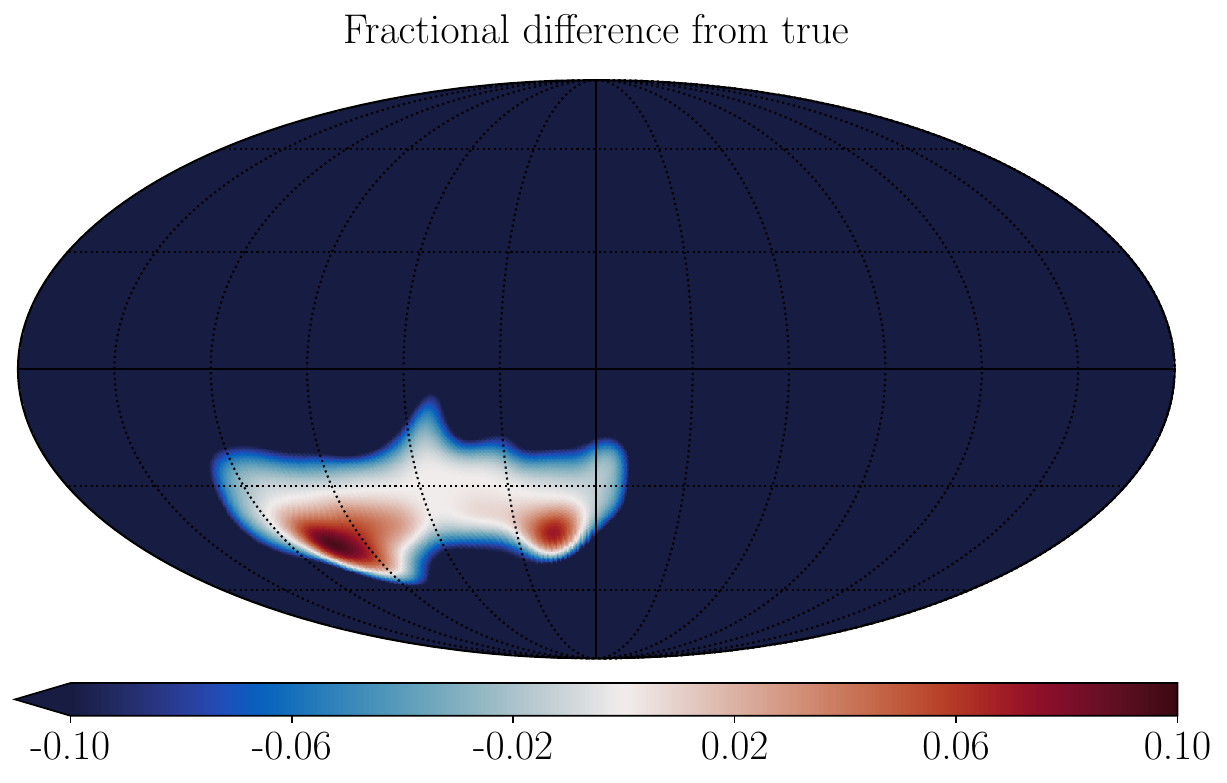}\hspace{0.06\columnwidth}
	\includegraphics[width=0.95\columnwidth]{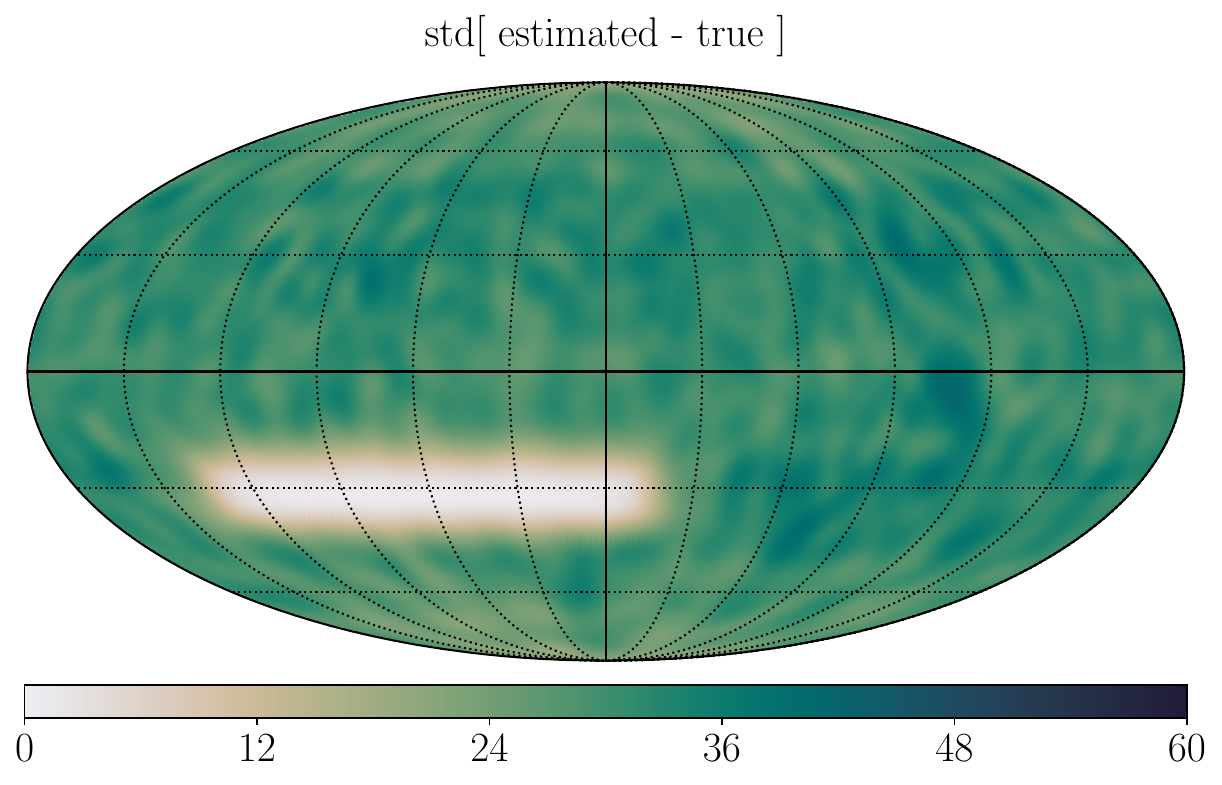}
    \caption{Maps generated from the estimated spherical harmonic modes on the sky using the \emph{standard} configuration and parameters. \emph{Upper left:} The true sky given by \texttt{pyGSM} with $\lmax=20$ and \nside$=128$. Note that the rippled structure comes from the true sky and not from the GCR solver. \emph{Upper right:} The estimated sky based on the mean of \comment{200} samples from the GCR solver. The spherical harmonic modes can be seen in Fig.~\ref{fig:alms_standard}. \emph{Bottom left:} Fractional difference between the mean and the true sky  adjusted to show differences $<10\%$, which coincides with the beam region, see also Fig.~\ref{fig:cart_beam_residual} for a closeup of the region. \emph{Bottom right:} The standard deviation of the difference of the estimated and true sky.}
    \label{fig:sky_maps}
\end{figure*}

\section{Results}
\label{sec:results}
In the following we show results of the samples obtained under various different parameter configurations of the sampler itself and by investigating the effects of different observation scenarios. 
The purpose of this analysis is to demonstrate the basic behaviours of the method and to validate that it can recover the true sky to a reasonable level. The comparative analysis uses simulations in order to understand the behaviour we would expect from real data as a function of different LST coverage, the effects of effectively truncating the visibility response operator at different levels of \lmax, and the impact of increasing the FoV of the array. We also dive into a stress-test of the prior- vs likelihood levels, making sure that neither is too broad nor too narrow.

Before going into the comparative analysis, we show in more detail in Sec.~\ref{subsec:the_standard_case} the results of our chosen reference -- or, \emph{standard} -- case, using the general parameters described in Sec.~\ref{sec:simulations}, considering not only the fractional difference between the posterior mean and true sky map but also looking at the recovered \alm modes, followed by the full comparative analysis in Secs.~\ref{subsec:noise_and_prior_levels}--\ref{subsec:array_and_observing_configs}.
Unless otherwise stated, we draw \comment{200} samples per scenario, as we deemed this sufficient for calculating the relevant statistics of the recovered spherical harmonic modes.
As a precautionary check of this assumption, we also consider a high-sample size case of $N_\text{samples}=5000$.
Finally, Sec.~\ref{subsec:beam_3_m_mwa_like} is dedicated to a closer study of the special case of increasing the field of view by decreasing the diameter of the antennas to $\theta_\text{D}=\SI{3}{\meter}$, which results in a FWHM at $\SI{100}{\mega\hertz}$ of $\sim \SI{60}{\degree}$. Having a greater sky coverage should improve on the sensitivity on lower $\ell$ modes. A large FoV can be achieved with instruments like the MWA.  

\begin{figure*}
	\includegraphics[width=\textwidth]{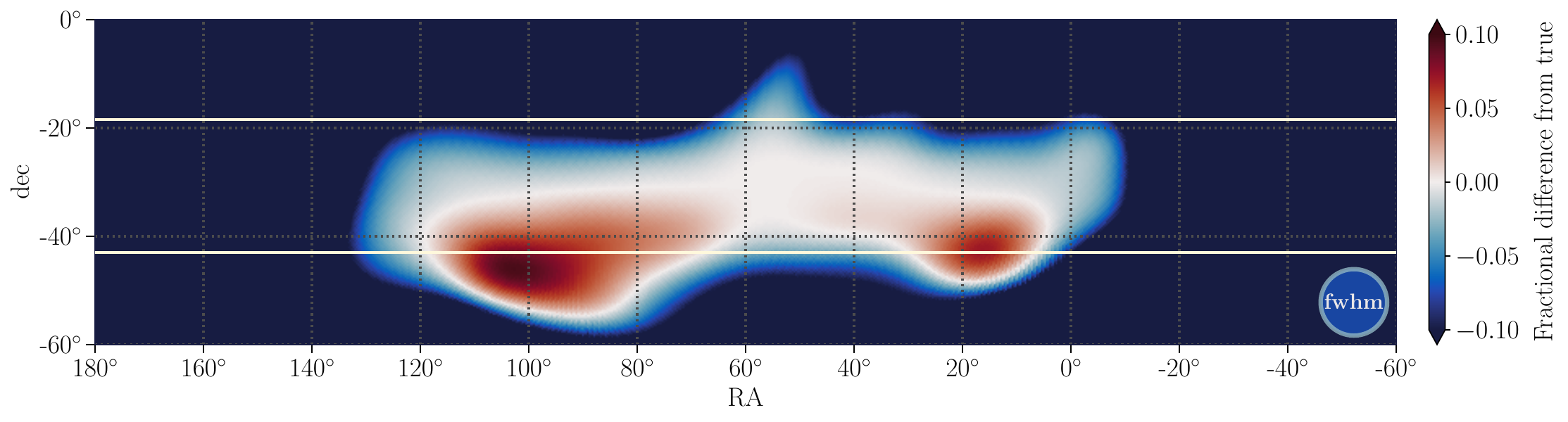}
    \caption{Cartesian projection of the fractional difference from the true sky using the \emph{standard} configuration and parameters. The RA and dec ranges have been narrowed to focus on the primary beam region. The FWHM of \SI{12.3}{\degree} is illustrated in the corner of the figure. The white horizontal lines indicate a FWHM distance from the centre of the primary beam at \SI{-31.7}{\degree}. The simulated LST range is \num{0}-\SI{8}{\hour} corresponding to an RA range of $\SI{0}{\degree}$-$\SI{120}{\degree}$. The centre of the beam-region has fractional differences  to the true sky of $<5\%$ and at the bounds $<10\%$.}
    \label{fig:cart_beam_residual}
\end{figure*}
\begin{figure}
	\includegraphics[width=\columnwidth]{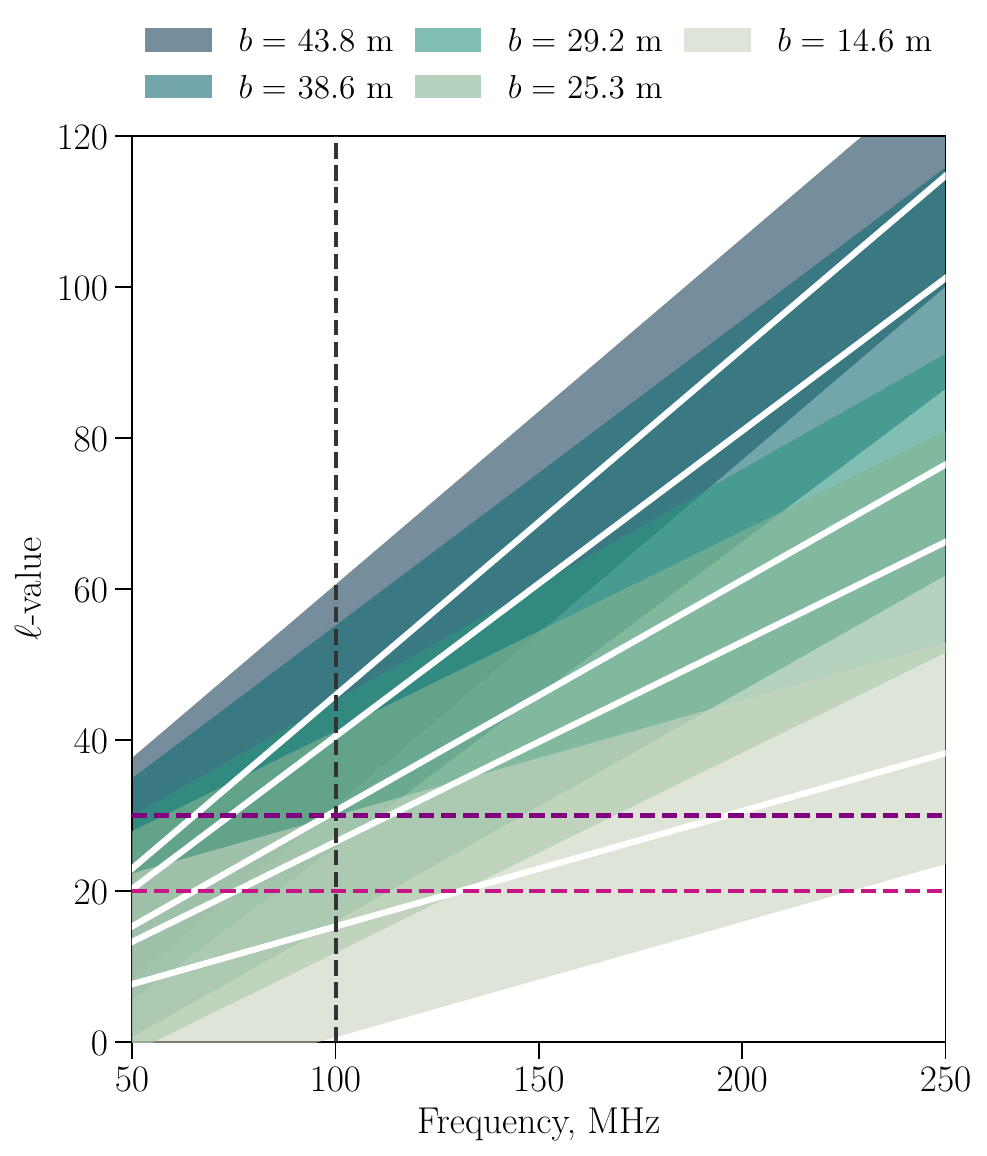}
    \caption{Approximate range of $\ell$-value sensitivity \comment{($\ell_\textup{min}\sim (\pi b)/\lambda$)} for several baselines, as a function of frequency (white lines with colour-coded shaded regions). The baselines displayed here are the five unique baseline lengths of the 10-dish standard configuration. The width (shaded regions) is given by the FWHM of the beam $\delta\ell\simeq\pi/{\rm FWHM}$, shown here for the HERA-like case with FWHM $=\SI{12.3}{\degree}$. The two horizontal lines indicate an $\lmax=30$ (top, dark purple) and $\lmax=20$ (bottom, magenta) and the black vertical line is set at $\nu=\SI{100}{\mega\hertz}$, the frequency used for the simulations in this work.}
    \label{fig:baseline_vs_ell}
\end{figure}

\subsection{realizations of the standard configuration}
\label{subsec:the_standard_case}
\begin{figure*}
	\includegraphics[width=0.98\textwidth]{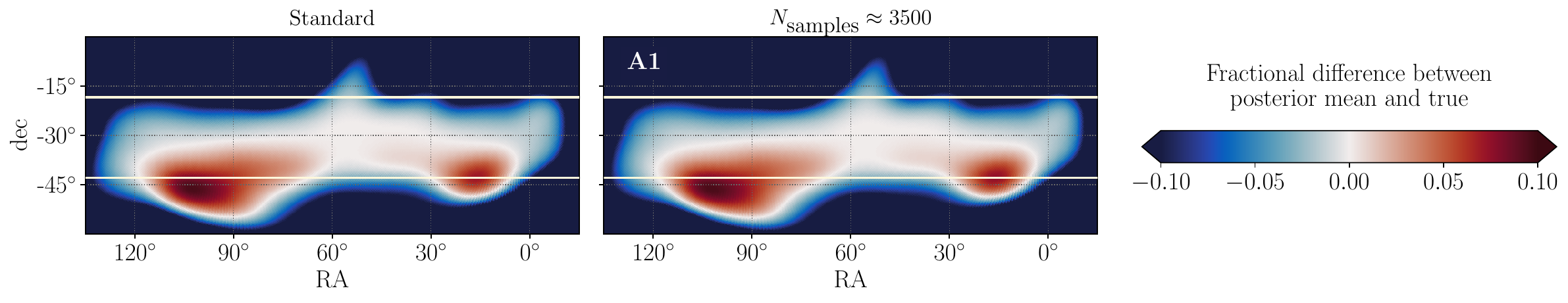}
 	\includegraphics[width=0.98\textwidth]{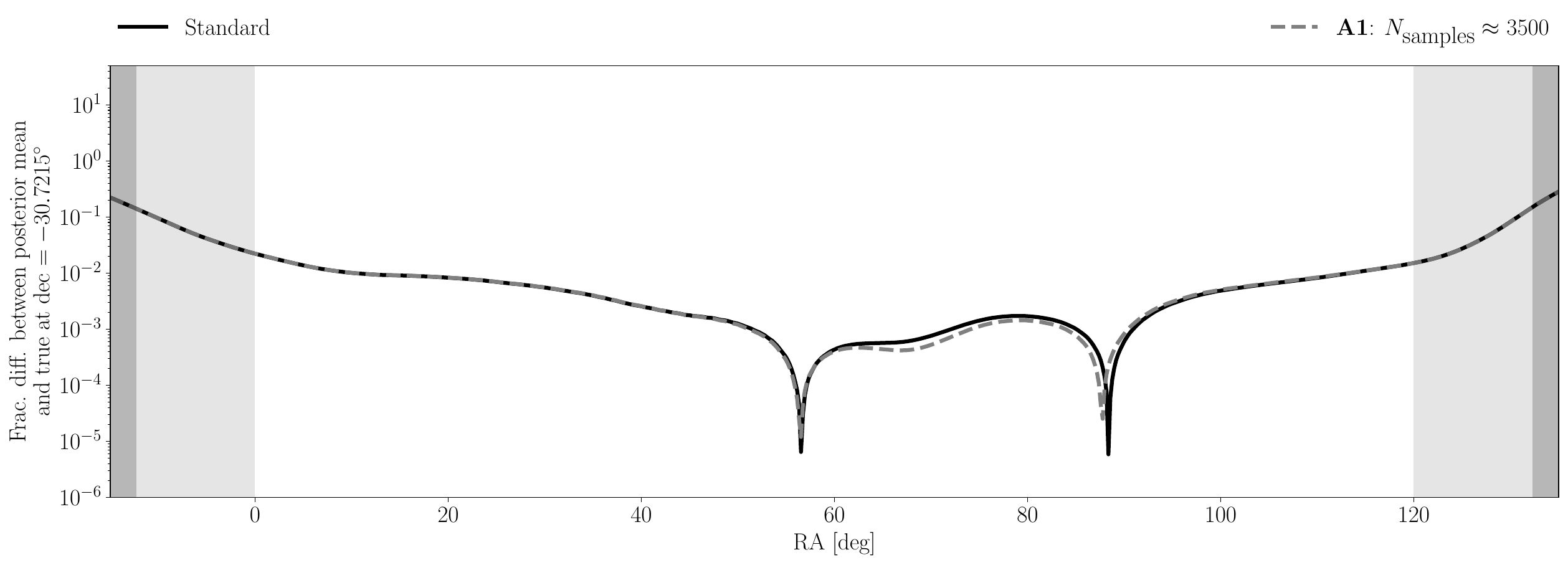}
    \caption{\comment{\emph{Top:} Cartesian projection of the fractional difference between the posterior mean and true sky for a high-sample size case (\case{A1}) compared to the \emph{standard} case. The fractional difference is defined as defined as $(\mu({\vecb{a}}_\textup{cr})-\vecb{a}_\textup{true})/\vecb{a}_\textup{true}$, where $\mu$ indicates the sample mean. The two horizontal lines indicate a FWHM distance from the centre of the beam. \emph{Bottom:} The absolute value of the fractional difference between the posterior mean and true sky taken as a slice through the centre of the primary beam region at dec=$\SI{-31.7}{\degree}$.  Artefacts due to the pixelisation have been smoothed away by applying a pair of moving average filters to each of these curves. For all runs in this plot the LST range is $0-\SI{8}{\hour}$ corresponding to RA values of $0-\SI{120}{\degree}$. The light shaded region covers the FWHM of $\SI{12.3}{\degree}$ and the dark shaded region is outside of the primary beam.}}
    \label{fig:cartesian_slice_nsample}
\end{figure*}

The results from the GCR of the \alm modes for the standard case are shown in Figs.~\ref{fig:alms_standard},~\ref{fig:sky_maps}, and \ref{fig:cart_beam_residual}. First, we define the true-sky subtracted mean as the mean of the spherical harmonic coefficients sampled by the GCR method subtracted by the true sky (input model) spherical harmonic modes from \texttt{pyGDSM}. In Fig.~\ref{fig:alms_standard} the true-sky subtracted mean of \comment{200} realizations of the \alm modes is shown normalised by the standard deviations of the samples. To be able to show the details of the region around a difference of zero, the figure is cropped to $\pm15\sigma$ with any modes outside of this region are marked with a $\blacktriangle$ and their values. Since the $m=0$ imaginary modes are artificially injected back in with a value of 0 (cf. Eq.~\ref{eq:real-field_anti_symmetry_condition}), they are without error bars but have been included and marked with $\times$ in the figure for clarity. It is clear that the higher order modes are easiest to constrain and that the lower order modes are far from the true sky.

In Fig.~\ref{fig:baseline_vs_ell} the $\ell$-mode sensitivity given the baseline length and frequency is shown. The shortest, and most frequent, baseline length in the standard configuration is $b=\SI{14.6}{\meter}$ corresponding to an $\ell_\textup{min}\sim (\pi b)/\lambda\sim 15$ at a frequency of $\SI{100}{\mega\hertz}$. Note, that at this frequency the baseline length (in metres) roughly corresponds to the $\ell$-value. Thus we are not expecting to be sensitive to $\ell\sim$~few, since modes with $\Delta \theta \gtrsim \lambda / b_{\rm min}$ are resolved out by the interferometer, and so we should not expect the data to constrain them. The recovered values of these low-$\ell$ modes are instead prior-driven.

In fact, Fig.~\ref{fig:alms_standard} is showing the spherical harmonic coefficients, each of which integrates information across the whole sky. Since the observed region is small (only $\sim$ 8.6\% of the sky), the recovered \alm values are generally mostly determined by the prior even when the noise level is very low in the observed region, we suspect this is because the prior-dominated area is much larger. Hence, plotting the statistics of the recovered \alm values directly is not a very sensitive test of this method.

Instead, it is much more useful to look at the recovered sky in map-space, as the difference between the observed and unobserved region is much clearer laid out.
In Fig.~\ref{fig:sky_maps} we show the maps of the true input sky (top left) given as described in Sec.~\ref{subsec:diffuse_emission_sky_model_pygsm} compared to the recovered map defined by the mean of the posterior (top right). It is noted that the recovered map shows lower values for the large scale modes (just as the $\alm$-values in Fig.~\ref{fig:alms_standard}) compared to the true sky map, which we expect is due to a random fluctuation rather than a bias.

Fig.~\ref{fig:sky_maps} also shows the fractional difference between these two maps (bottom left) and the standard deviation of the recovered sky (bottom right). The fractional difference between the posterior mean and true sky is lowest within the primary beam region and the standard deviation is much smaller in this area too, thus demonstrating the specific patch of sky that is directly observed versus that constrained by the prior. This explains how we are still able to recover most of the galactic structure, despite the majority of it being outside the observed region of the sky.

Now focusing on the primary beam region alone, we show a Cartesian projection of the fractional difference between the posterior mean and true sky in Fig.~\ref{fig:cart_beam_residual}. The region is defined to be within a FWHM on either side of the central latitude of \SI{-31.7}{\degree}. For the standard case with the antenna diameter of $\theta_\text{D}=\SI{14}{\meter}$ the FWHM is \SI{12.3}{\degree}. For the standard case the sky is recovered to within 5\% close to the centre of the beam region and to within 10\% within the FWHM bounds. 

\begin{figure*}
	\includegraphics[width=0.95\textwidth]{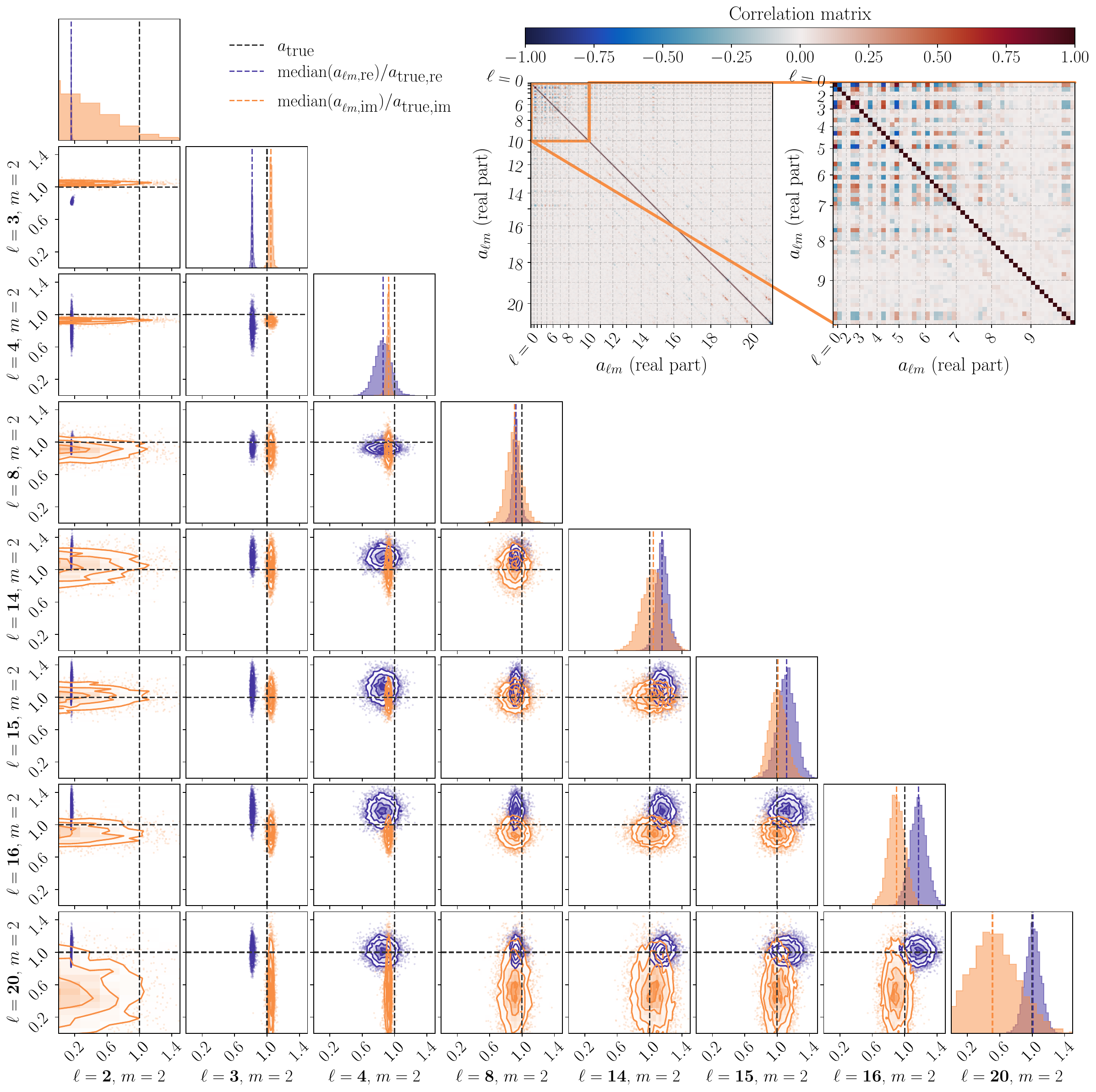}
    \caption{\comment{\emph{Top right:} Correlation matrix for the high-sample size \case{A1}-case (real part) with an extra panel zoomed in on the $\ell\leq9$ modes. \emph{Bottom left:} The marginal 1D- and 2D-posterior distributions normalised by the true values for the high-sample \case{A1}-case for a range of $\ell$ modes at $m=2$. The $\ell$-modes have been chosen to have all scales represented. The median of the marginal posteriors for the real- and imaginary part is shown in colour along with the true value (black). The $m$-value is chosen at random and held the same for consistency.}}\label{fig:corner_plot_correlation_mat}
\end{figure*}

\begin{figure*}
	\includegraphics[width=\textwidth]{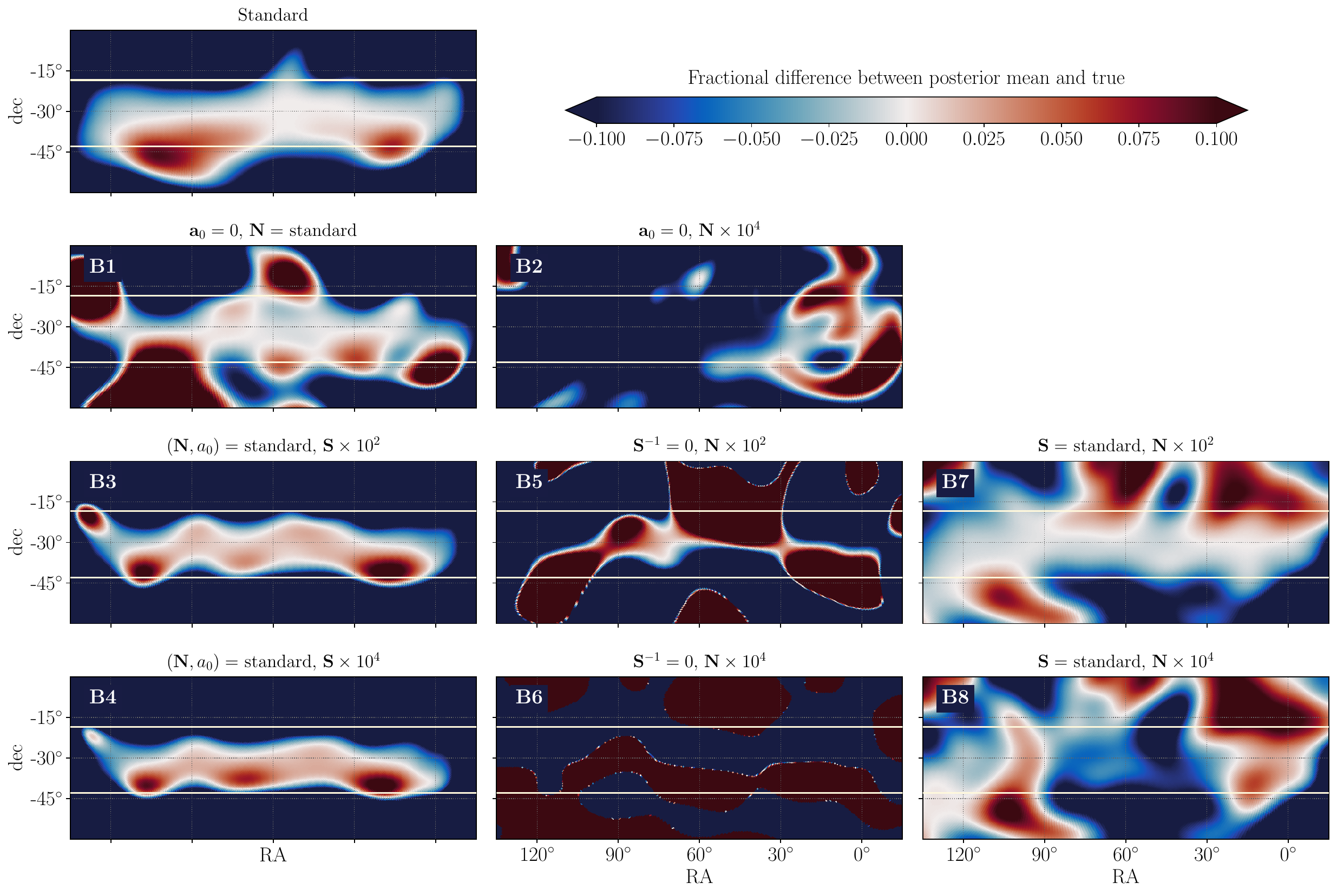}
    \caption{\comment{Cartesian projection of the fractional difference between the posterior mean and true sky for multiple sets of samples with varied levels of noise covariance $\vecb{N}$, prior covariance $\vecb{S}$, and prior mean $\vecb{a}_0$ compared to the \emph{standard} case.} The fractional difference is defined as defined as $(\mu({\vecb{a}}_\textup{cr})-\vecb{a}_\textup{true})/\vecb{a}_\textup{true}$, where $\mu$ indicates the sample mean. The two horizontal lines indicate a FWHM distance from the centre of the beam.}
    \label{fig:cartesian_comparison_noise_prior}
\end{figure*}

\addtolength{\tabcolsep}{+5.2pt}   
\begin{table}
    \centering
        \caption{Overview of all the cases analysed in this paper and their labels along with which parameters are varied. A ``---'' indicates that the parameter is set as in the standard case. The labels are numbered after the order of appearance in \comment{Figs.~\ref{fig:cartesian_slice_nsample},~\ref{fig:cartesian_slice_noise_prior}~and~\ref{fig:cartesian_slice_observ}}. }
    \begin{tabular}{c c c c}
    \toprule
     & $N_{\rm samples}$ & & \\
    \midrule
    \case{A1} & \comment{$5000$} & --- & --- \\
    \midrule
         & Noise covariance & Prior covariance & Prior mean  \\ 
    \midrule
    \comment{\case{B1}} & --- & --- & $\vecb{a}_0 = 0$ \\
    \comment{\case{B2}} & $\vecb{N}\times10^4$ & --- & $\vecb{a}_0 = 0$ \\
    \comment{\case{B3}} & --- & $\vecb{S}\times10^2$ & --- \\
    \comment{\case{B4}} & --- & $\vecb{S}\times10^4$ & --- \\ 
    \comment{\case{B5}} & $\vecb{N}\times10^2$ & $\vecb{S}^{-1}=0$ & --- \\
    \comment{\case{B6}} & $\vecb{N}\times10^4 $ & $\vecb{S}^{-1}=0$ & --- \\
    \comment{\case{B7}} & $\vecb{N}\times10^2$ & --- & --- \\
    \comment{\case{B8}} & $\vecb{N}\times10^4$ & --- &  --- \\
    \midrule 
        & $\lmax $ & Number of LSTs & LST range  \\
    \midrule
    \comment{\case{C1}} & $\lmax=30$ & --- & --- \\
    \comment{\case{C2}} & --- & $N_{\rm LST}=10$ & $8-\SI{16}{\hour}$  \\
    \comment{\case{C3}} & --- & $N_{\rm LST}=10$ & $16-\SI{24}{\hour}\,\,\,$  \\
    \comment{\case{C4}} & --- & $N_{\rm LST}=20$ & $0-\SI{8}{\hour}\,\,\,$\\
    \comment{\case{C5}} & --- & $N_{\rm LST}=20$ & $0-\SI{16}{\hour}$ \\
    \midrule
     & $N_{\rm ants}$ & Dish diameter & FWHM \\
    \midrule
    \comment{\case{D1}} & --- & $\SI{3}{\meter}$ & $\SI{57.3}{\degree}$ \\
    \bottomrule
    
    \end{tabular}
    \label{tab:overview_of_all_run_labels}
\end{table}
\addtolength{\tabcolsep}{-5.2pt}   

\comment{
\subsection{High number of samples, case \case{A1}}}
\comment{
\noindent For the standard case we deliberately keep the sample size low to keep down the computational cost. However, an additional run was made with $N_{\rm samples}\comment{= 5000}$ (dubbed case~\case{A1} (see Table~\ref{tab:overview_of_all_run_labels} for a full overview of all cases) and we examine whether using many more samples would significantly change the results, in order to check that only \comment{200} samples is enough to produce statistically valid results.
The fractional difference between the posterior mean and true sky for the \case{A1}  result is shown in Fig.~\ref{fig:cartesian_slice_nsample} (top) along with re-displaying the standard case for easier comparison. It is clear that the fractional difference of the two cases look almost identical. 

For even closer comparison we take a slice of the centre of the beam of the absolute values of the fractional difference to the true value, as shown in Fig.~\ref{fig:cartesian_slice_nsample} (bottom). As the figure shows, even when comparing the fractional difference slices, the two cases only have a slight difference at ${\rm RA} \sim \SI{10}{\degree}$, thus emphasising that these two results are almost identical, which (together with the full beam region fractional difference) justifies the use of \comment{200} samples as adequate for our purposes.
}

\comment{
Before diving into the next test cases, where the noise and prior levels are varied, we look at the high sample case (\case{A1}) in more detail. 
In Fig.~\ref{fig:corner_plot_correlation_mat} (top right) we show the correlation matrix for the real part of the spherical harmonic modes. The imaginary part has been left out, solely to more clearly display the matrix. The lower $\ell$-modes are generally more correlated, which makes sense since they are larger than the FoV. A zoom-in of these modes is also displayed to better see the hatched pattern their correlations form. We believe the pattern is due to variations in sensitivity to specific modes due to baseline orientation as well as general lower sensitivity (and therefore easier mixing of) the lower $\ell$ modes. It is also noticeable that (mainly visible for the larger $\ell$-modes) there are diagonal structures outside of the main diagonal. This implies that neighbouring modes are slightly correlated.
}

\comment{
Because the sensitivity to the low-$\ell$ modes is lower and they are more correlated, we expected this to result in broader constraints on the low-$\ell$ modes. However, specifically what we see from the posterior distributions for the real-part (purple) in Fig.~\ref{fig:corner_plot_correlation_mat} is that the lower modes are biased rather than poorly constrained. The imaginary-part, on the other hand, indeed shows broader contours both at low $\ell$ and for $\ell=20$. Outside of the low-$\ell$ modes, the results are fairly non-biased --- especially the modes around the peak-sensitivity ($\ell=14, 15$, and $16$; Fig.~\ref{fig:baseline_vs_ell} demonstrates baseline and $\ell$ relation). 
All in all, higher-$\ell$ modes are well constrained, and are statistically consistent with the input value, whereas the very largest modes (low-$\ell$), corresponding to scales larger than the field of view, have smaller error-bars than expected, and hence look better constrained than they really are.
}

\subsection{Noise and prior levels}
\label{subsec:noise_and_prior_levels}
\begin{figure*}
	\includegraphics[width=\textwidth]{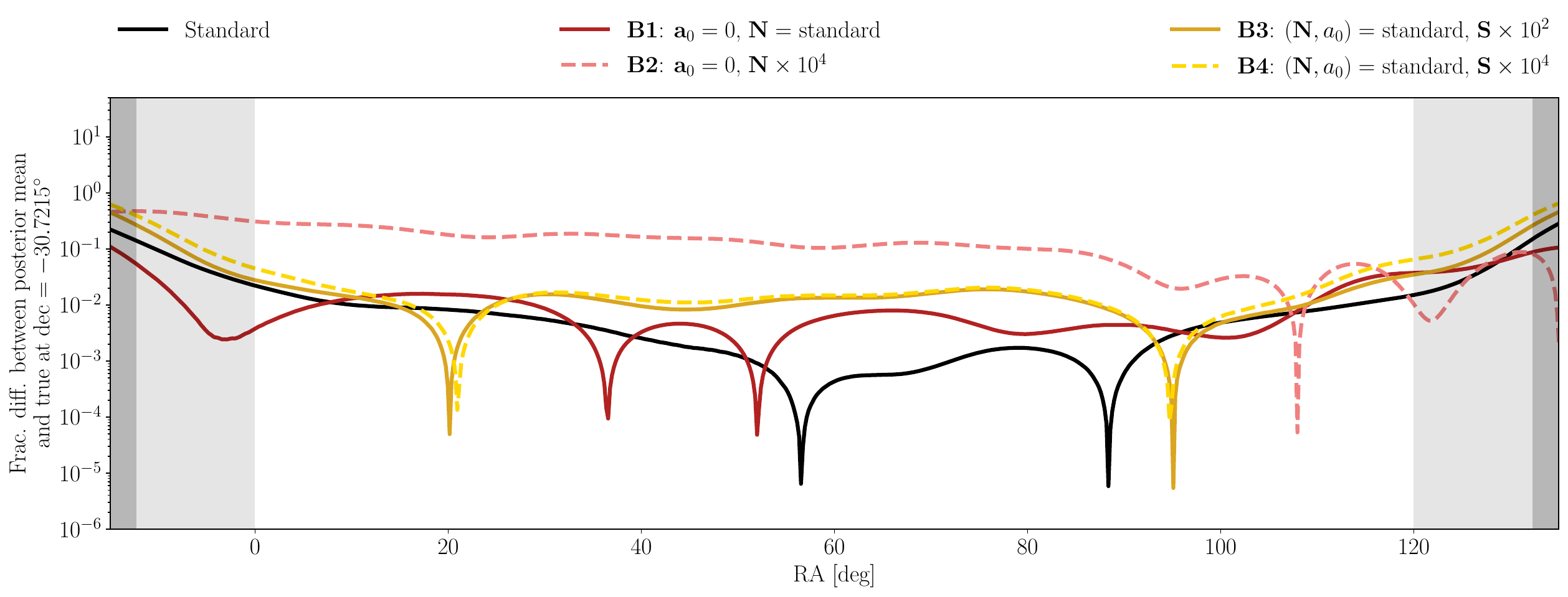}
	\includegraphics[width=\textwidth]{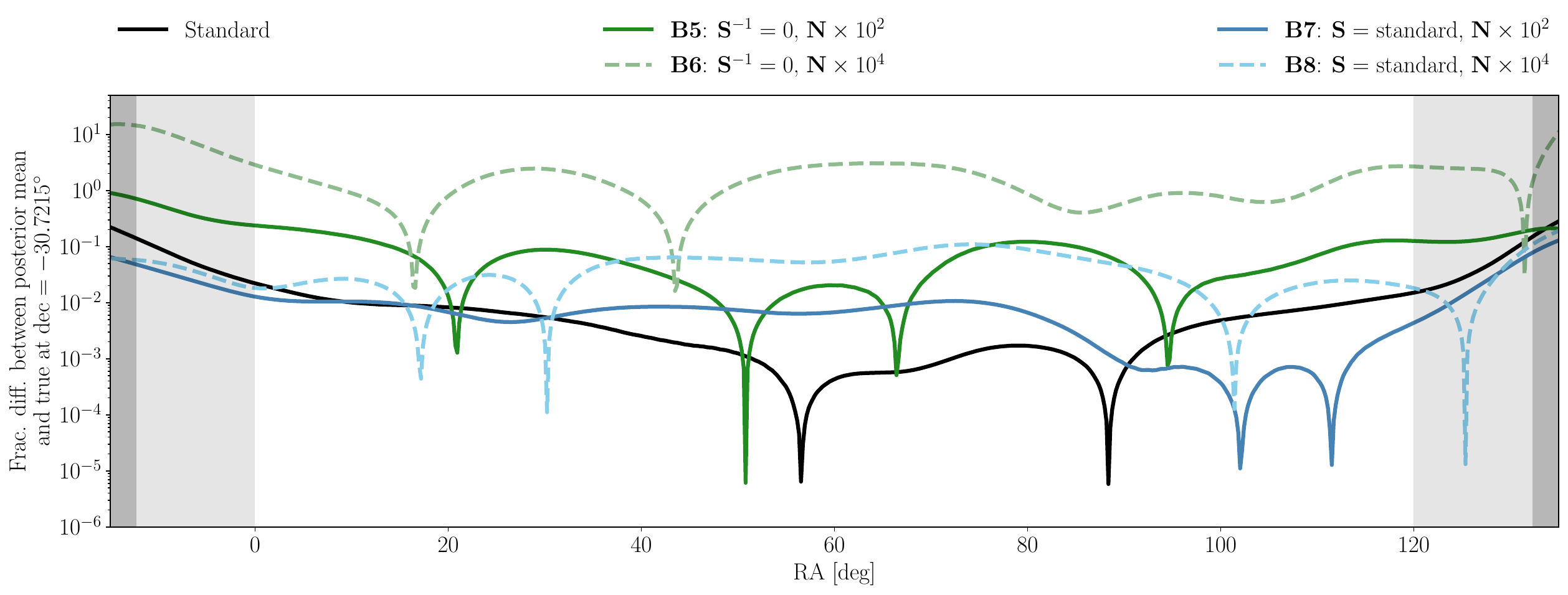}
    \caption{The absolute value of the fractional difference between the posterior mean and true sky taken as a slice through the centre of the primary beam region at dec=$\SI{-31.7}{\degree}$.  Artefacts due to the pixelisation have been smoothed away by applying a pair of moving average filters to each of these curves. For all runs in this plot the LST range is $0-\SI{8}{\hour}$ corresponding to RA values of $0-\SI{120}{\degree}$. The light shaded region covers the FWHM of $\SI{12.3}{\degree}$ and the dark shaded region is outside of the primary beam.}
    \label{fig:cartesian_slice_noise_prior}
\end{figure*}
\begin{figure*} 
	\includegraphics[width=\textwidth]{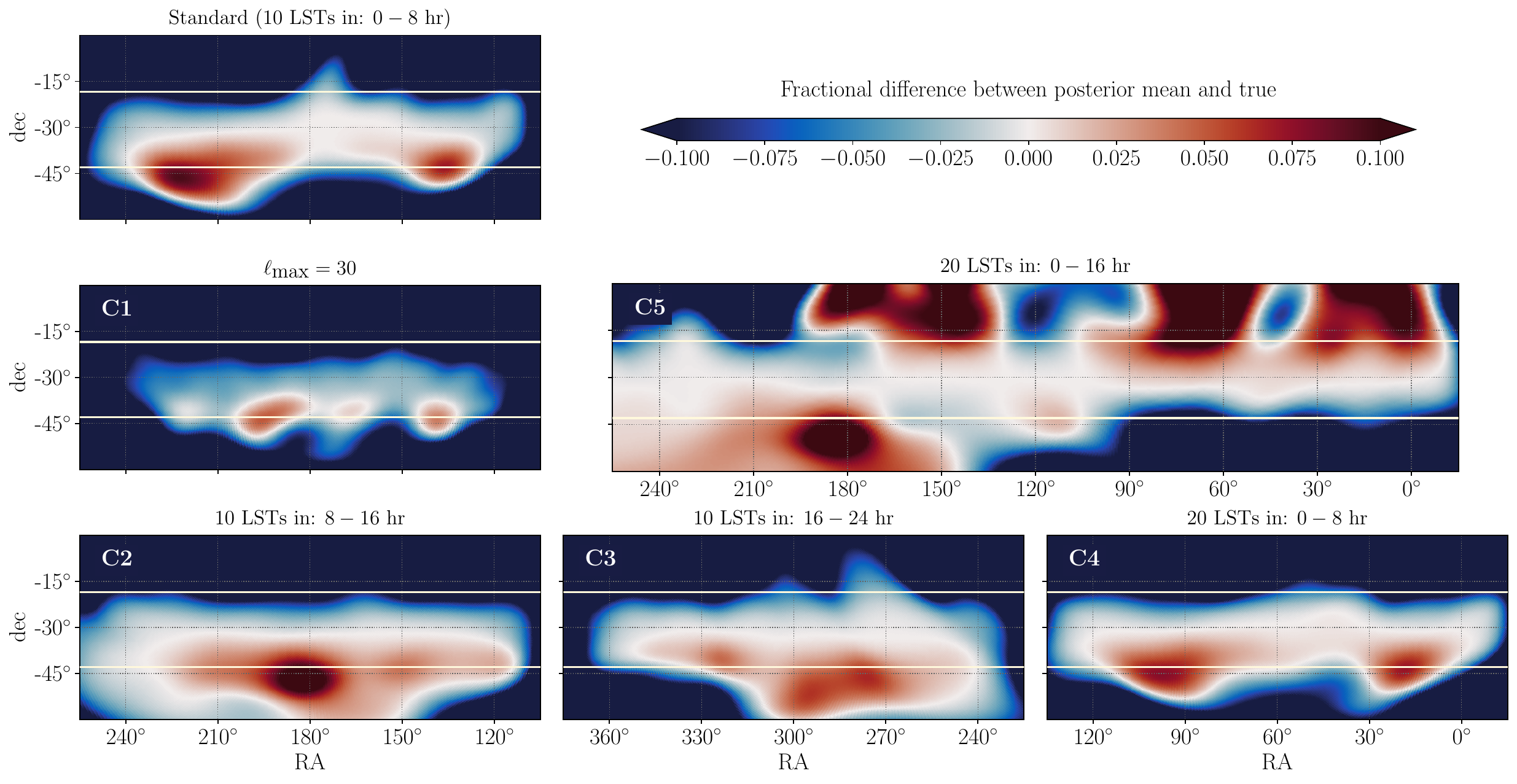}
    \caption{Cartesian projection of the fractional difference between the posterior mean and true sky for runs with variations to the standard observation/simulation parameters such as increasing the $\lmax$, adding antennas for longer baselines, increasing the LST range, and increasing the cadence of observations. The fractional difference is defined as defined as $(\mu({\vecb{a}}_\textup{cr})-\vecb{a}_\textup{true})/\vecb{a}_\textup{true}$, where $\mu$ indicates the sample mean. The two horizontal lines indicate a FWHM distance from the centre of the beam.}
    \label{fig:cartesian_comparison_observing_configs}
\end{figure*}

\begin{figure*}
	\includegraphics[width=\textwidth]{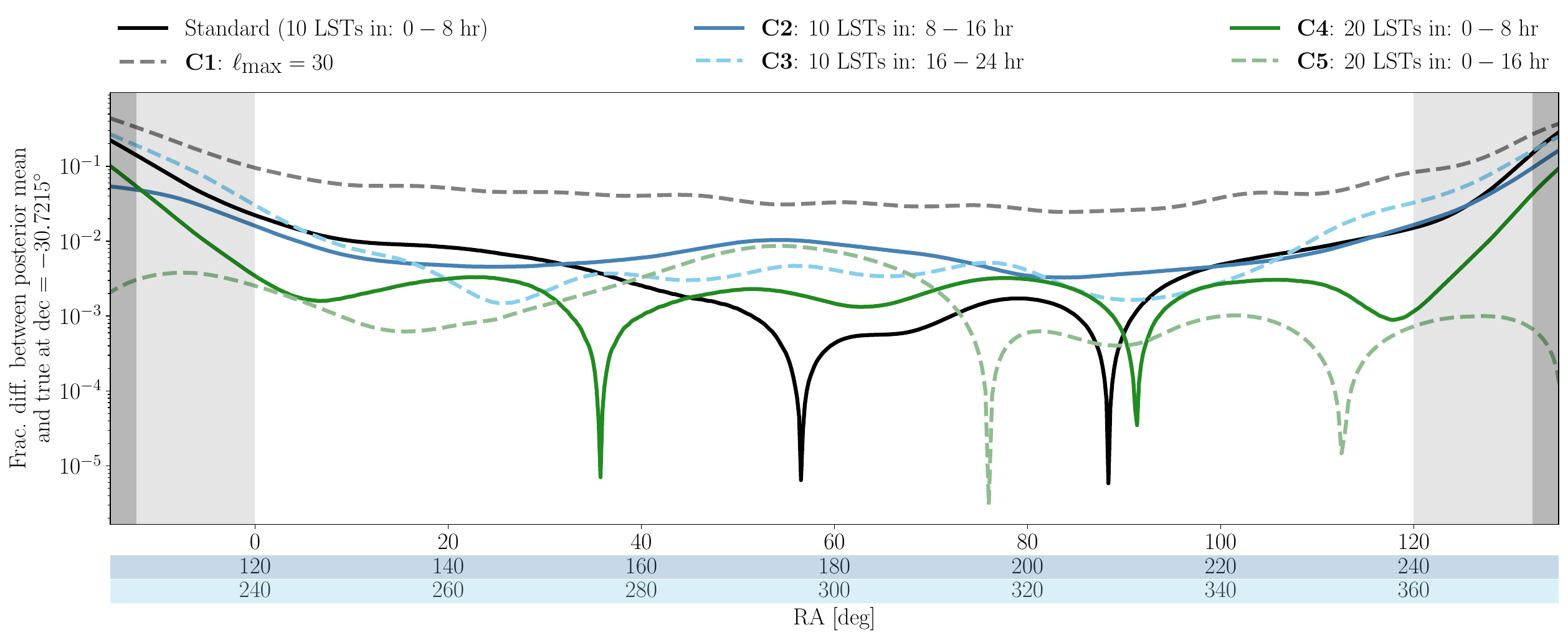}
    \caption{The absolute value of the fractional difference between the posterior mean and true sky taken as a slice through the centre of the primary beam region at dec=$\SI{-31.7}{\degree}$. Artefacts due to the pixelisation have been smoothed away by applying a pair of moving average filters to each of these curves. \comment{Results~{\case{C3}-\case{C2}}} have different LST/RA ranges as indicated by the additional x-axes. Note that for \case{C5}, only the first half of the LST/RA range is shown, to match the $\SI{8}{\hour}$ span of the other runs. The light shaded region covers the FWHM of $\SI{12.3}{\degree}$ and the dark shaded region is outside of the primary beam. It is noticeable that all three $\lmax=30$ results have higher fractional differences to the true than the $\lmax=20$ results.}
    \label{fig:cartesian_slice_observ}
\end{figure*}

In this section we explore how varying the noise covariance $\mathbf{N}$, prior covariance $\mathbf{S}$, and prior mean $\vecb{a}_0$ will affect the results from the GCR-sampler compared to our chosen standard case. The motivation for this is to assure that a sufficient balance is reached between the data- and prior-information, and that the prior is not dominating the system. The results of the various noise- and prior level runs can be seen in Figs.~\ref{fig:cartesian_comparison_noise_prior} and \ref{fig:cartesian_slice_noise_prior}. All the various runs have been labelled in order of \comment{ their} appearance in Fig.~\ref{fig:cartesian_comparison_noise_prior} and these labels will be used throughout the paper. A full overview of all labels can also be seen in Table~\ref{tab:overview_of_all_run_labels}.

An important test is to make sure that our choice of prior mean, $\vecb{a}_0$, is not over-informing the sampler. Comparing the \case{B1} ($\mathbf{a}_0=0, \mathbf{N}=$ standard) and \case{B2} ($\mathbf{a}_0=0, \mathbf{N}\times 10^4$) cases to the standard case in  Fig.~\ref{fig:cartesian_comparison_noise_prior}, it initially looks like setting the prior mean to zero results in a poorly recovered sky. Indeed, this is the case if the noise is also increased (as in \case{B2}), although generally increasing the noise with a factor of $10^4$ has yielded poor results (\case{B4}, \case{B6}, \case{B8}) and does not seem to be due to the change of the prior mean alone. Instead, when the noise level is kept at standard level (case \case{B1}), the true sky is still well-recovered near the centre of the beam (i.e. recover is not biased), but the residual is larger beyond the beam FWHM than in the standard case. Again, taking a slice of the absolute fractional difference at the central declination only (around $\textup{dec}=\SI{-31.7}{\degree}$), Fig.~\ref{fig:cartesian_slice_noise_prior}), shows that the two $\mathbf{a}_0=0$ cases are not performing as good as the standard case. Although, especially for the \case{B1} case, at RA$=60-\SI{100}{\degree}$ it is performing similarly or (very) slightly better than the standard case. In most of our runs we have a 10\% prior that has prior means which are scattered around the true value. This is somewhat conservative (we are not assuming a 'correct' prior mean and thereby not enforcing the true solution), but there is still the question of how sensitive the results are to prior choices. What the $\vecb{a}_0=0$ runs (\case{B1}, \case{B2}) show is what happens when our prior is typically wrong by $10\sigma$.

\comment{Next,} we set the inverse prior covariance to zero, which may be interpreted as the case where only the likelihood term is used and the prior is neglected. We then multiply the noise covariance with a factor of $10^2$ and $10^4$, case~\case{B5} and \case{B6} respectively. This is done in order to confirm, that removing the information provided by the likelihood and/or prior does indeed lead to no recovery of information of the sky, i.e. we demonstrate that the good recovery in other cases is not simply due to the structure of the sky model or similar. The fractional difference between the posterior mean and true sky for the \case{B5} case ($\vecb{S}^{-1}=0, \mathbf{N}\times10^2$) shows a slight hint of the primary beam structure, but the high noise levels washes out the data and there is no prior-information to assist the recovery of the sky. When increasing the noise further in the \case{B6} case ($\vecb{S}^{-1}=0, \mathbf{N}\times10^4$), the recovered sky is all gone and there is only noise left. In order to see how well the sampler can recover the sky with no prior information, an extra run (not shown) was made with standard noise level (but keeping the inverse prior covariance at zero)\comment{, i.e. equivalent to a pure maximum likelihood solution with a flat prior. This run} resulted in a fractional difference between the posterior mean and true sky very similar to that of \comment{ case~\case{B4}} (i.e. \comment{with} $\mathbf{S}\times10^4$). Essentially the very high value of the prior covariance $\mathbf{S}$ is the same as saying the inverse prior covariance $\vecb{S}^{-1}$ is closer to zero. It is therefore not that surprising that these two results would be similar. For both these cases the prior is now so broad and uncertain that the only well-recovered part is the observed region.  

The prior drives the good recovery outside of the observed region, which is very clear when comparing for instance case~\case{B8} ($\mathbf{S}=$ standard, $\mathbf{N}\times10^4$) to case~\case{B6} ($\mathbf{S}^{-1}=0, \mathbf{N}\times10^4$) \comment{in Fig.~\ref{fig:cartesian_comparison_noise_prior}}. Without any prior information nothing is recovered -- but with the prior most of the sky is recovered well to within $\sim10\%$. However, \case{B8} lacks the primary beam structure that we saw in the standard case, so the recovery here is due to the prior driving the solution since the broad noise covariance is washing out the data. When the noise is lowered again in case \case{B7} to $\mathbf{N}\times10^2$ (while keeping $\mathbf{S} =$ standard), the central beam structure starts to reappear. The good recovery of the sky outside of the observed region is without a doubt driven by the prior -- as the data cannot say anything about this region anyway -- but the \emph{very} good recovery of the observed region is due to the data. 
In the standard case the well-recovered sky is further limited to be close to the observed region. The fact that the spherical harmonics are continuous means the sky will still be constrained by the data a little outside of the observed region. This explains why the standard deviation in Fig.~\ref{fig:sky_maps} smoothly goes from low to a high value.

\subsection{Effect of changing array and observing configurations}
\label{subsec:array_and_observing_configs}
The performance of the sampler is affected by which spherical harmonic modes are available to it. This can be tested by changing the $\lmax$ or for instance by including longer baselines to increase sensitivity to higher $\ell$ modes, as can also be gauged from Fig.~\ref{fig:baseline_vs_ell}.
Since most of the diffuse foreground power comes from the Galaxy, it is also interesting to examine how changing the LST range will affect how well the sky is recovered. All the test cases in the following section has been labelled with a similar style to the previous section, now based on the order of appearance in Fig.~\ref{fig:cartesian_comparison_observing_configs}. A full overview of all labels can be seen in Table~\ref{tab:overview_of_all_run_labels}.

To capture regions of the entire HERA-strip we examine LST ranges of $8-\SI{16}{\hour}$ and $16-\SI{24}{\hour}$ on top of the $0-\SI{8}{\hour}$ LST range; case \case{C2}, \case{C3}, and the standard case, respectively. The \case{C3} case covers the brightest regions of the galactic diffuse emission, which can be seen in the RA range of $\SI{240}{\degree}-\SI{360}{\degree}$ (the readers right hand side) on the top-left plot of Fig.~\ref{fig:sky_maps}. We also examine the effects of increasing the data set to $N_\textup{LST}=20$ either by doubling the cadence and keeping the same LST range the same (case \case{C4}) -- or, vice versa (case \case{C5}). The former results in more information of the same sky region and the latter results in a greater sky coverage. The standard case cadence is one integration of duration 60 sec per 48 minutes equivalent of 10 samples evenly spread over an 8 hour LST range. Note, since the sky rotates $\SI{15}{\deg/\hour}$, the beam-crossing time of the HERA-like array is $\sim\SI{49}{\minute}$ given the FWHM of $\SI{12.3}{\degree}$. 
In this section we go over the results from each of these cases and how they affect the resulting recovered sky. 

For the standard case we use $\lmax = 20$ because it is a reasonable minimum for testing purposes that includes the shortest (most sensitive) HERA baseline, see Fig.~\ref{fig:baseline_vs_ell}.
In case~\case{C1}, we see from Fig.~\ref{fig:cartesian_slice_observ} that increasing $\lmax$ to 30 results in a significantly degraded fractional residual compared to $\lmax$ of 20.
While increasing $\lmax$ brings the $b=\SI{25.3}{\meter}$ baseline group into the \emph{directly} constrained $\ell$ range (see Fig.~\ref{fig:baseline_vs_ell}), each baseline is actually sensitive to a broad range of $\ell$-values, as for example the primary beam contributes to smearing of the $\ell$-mode sensitivity. For instance the $b=\SI{29.2}{\meter}$ baseline (antennas $(0,2)$) in Fig.~\ref{fig:vis_response_operator} is clearly sensitive to $\ell$-values down to around $\ell\sim8$ and the $(0,5)$-baseline, which is a shorter $\SI{25.3}{\meter}$ baseline, contributes mostly at $\ell\gtrsim5$ but still has some visibility response below this level. These baselines will therefore already contribute to some degree at the $\lmax=20$ level. However, when we increase the $\lmax$ to 30 the total number of modes more than doubles, from 441 to 961. This stretches the available signal-to-noise of the data much farther than in the $\lmax=20$ case.
\comment{It is noticable, however, that in map-space the fractional difference is relatively smooth for the $l_{\rm max}=30$ case  (\case{C1}, see Fig.~\ref{fig:cartesian_comparison_observing_configs}). Another possible explanation is that the solver is trying to fit the smaller scales modes to the monopole, and this is causing an overall amplitude difference.}

Next, we vary the LST ranges. First, we keep to 10 time steps and 8 hr ranges (as in the standard case), but shift the ranges to cover different parts of the sky. 
Fig.~\ref{fig:cartesian_comparison_observing_configs} shows that the range of $8-\SI{16}{\hour}$ (\case{C2}) performs better than both the standard $0-\SI{8}{\hour}$ and the \case{C3} case ($16-\SI{24}{\hour}$), which covers the brightest part of the galactic emission. As before, we inspect the slice at the central declination for a closer comparison, see Fig.~\ref{fig:cartesian_slice_observ}.
It is clear from the figure, that within their respective observed regions all three sets of LST ranges actually perform comparatively well with only small differences. As one moves closer to the edge of the observed region, however, the \case{C2} case shows some improvement to the standard case -- and, especially, when compared to the
\case{C3} case.
The difference in recovery level can simply be due to the shifting of the sampling points. For all the runs the simulated data is the same and have one specific realization of the noise. Areas that have less noise will therefore always have a higher signal-to-noise and should perform better. 

Increasing the data volume should also help constrain the sky better. One way to do this is to double the number of observation times from $N_\textup{LST}=10$ to $20$, either by  increasing the cadence within the same LST range (case~\case{C4}) or by 
doubling the range and keeping the cadence the same (case~\case{C5}). At first glance both results seem to perform similar or perhaps slightly better than the standard case, see Fig.~\ref{fig:cartesian_comparison_observing_configs}. 
The recovered signal will be worse at the edges of the observed region. The \case{C5} case is twice the size and encompasses the far edge of the \case{C4} case at LST $=\SI{8}{\hour}$, and will naturally be better recovered at this LST than the \case{C4} case. We will therefore limit the comparison to the $0-\SI{8}{\hour}$ range for the fractional residual slice in Fig.~\ref{fig:cartesian_slice_observ}. The \case{C4}  result mostly shows improvement to the standard case with the exception of RA $\SI{20}{\degree}-\SI{40}{\degree}$ although exceeded by the performance of the \case{C5} case, that not only performs better than the standard case but also remains closer to true at the lower edge ($\SI{0}{\hour}$ of the observed region.
This hints that increasing the observation time range (and thus increasing the size of the observable sky) has a larger impact on the recovery of the sky than doubling the data we have inside the same region. This can also be explained with a larger observable sky being sensitive to larger scales (lower spherical harmonic modes).

\subsection{Improving sensitivity to larger scales}
\label{subsec:beam_3_m_mwa_like}
\begin{figure*}  
	\includegraphics[width=\textwidth]{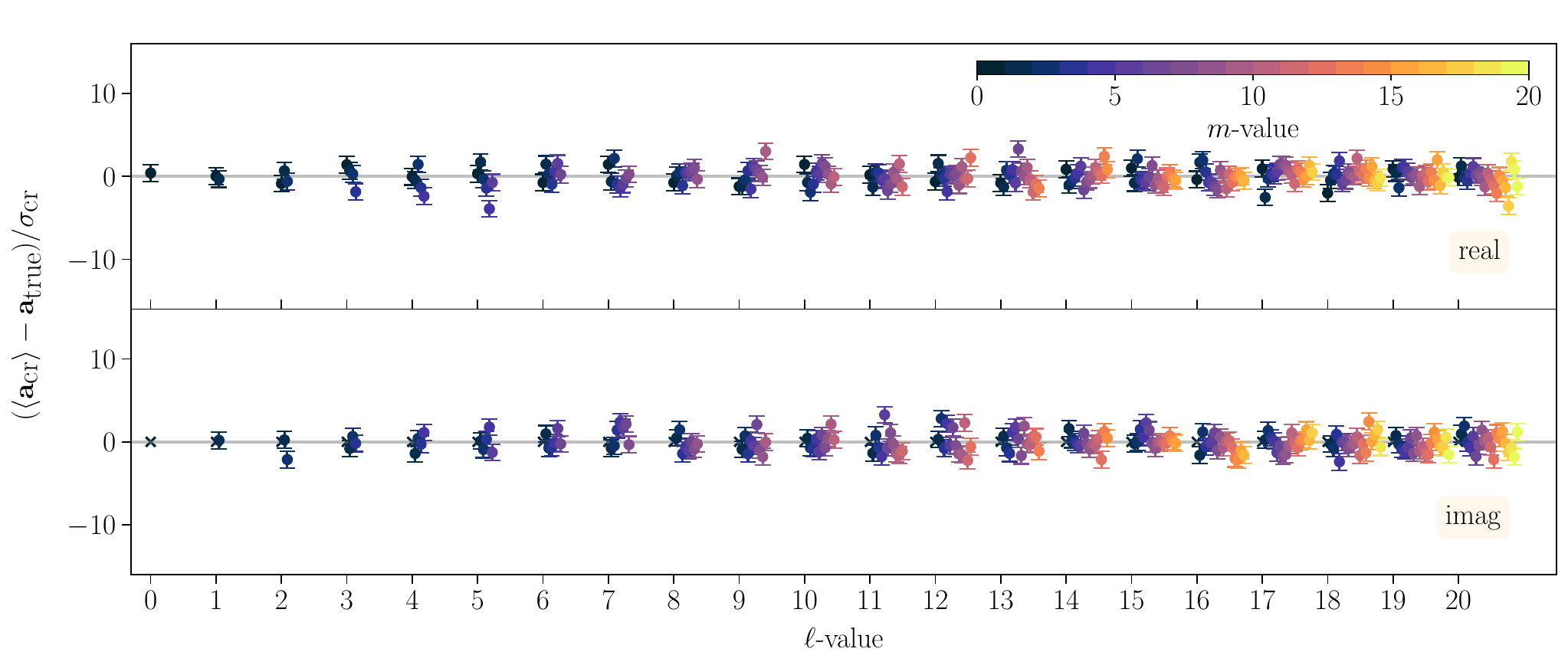}
    \caption{The real- and imaginary parts of the difference of the mean of \comment{200} samples of \alm modes (case \case{D1}) from the GCR solver and the true sky normalised by the individual standard deviations. Here, the diameter of the receivers set to $\theta_{D}=\SI{3}{\meter}$ and thereby increasing the field of view. As described in Sec.~\ref{subsec:data_model} the $m=0$ imaginary modes ($\times$) should always be zero and the GCR solver therefore does not solve for this. All residual $\alm$ modes now lie within the $\pm 15\sigma$ boundaries and including the uncertainties most modes are consistent with zero. } 
    \label{fig:alms_mwa}
\end{figure*}
\begin{figure*}
	\includegraphics[width=0.95\columnwidth]{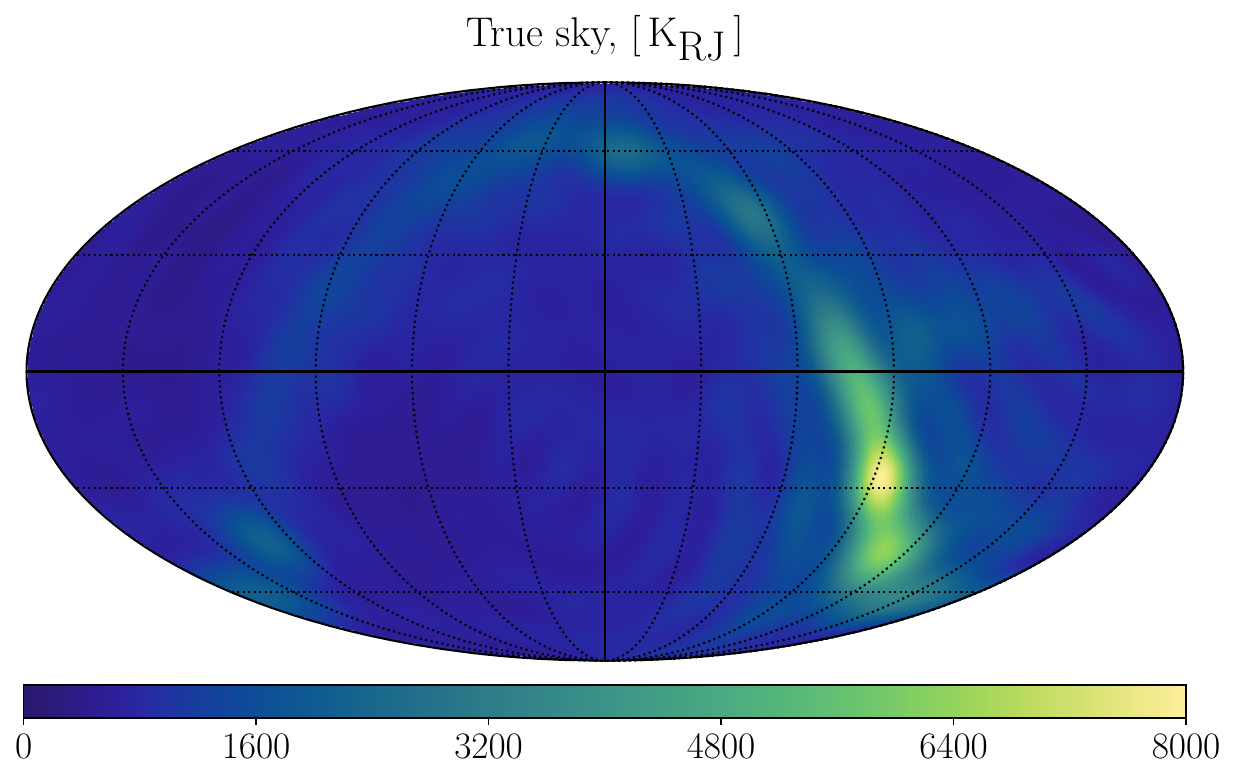}\hspace{0.06\columnwidth}
 	\includegraphics[width=0.95\columnwidth]{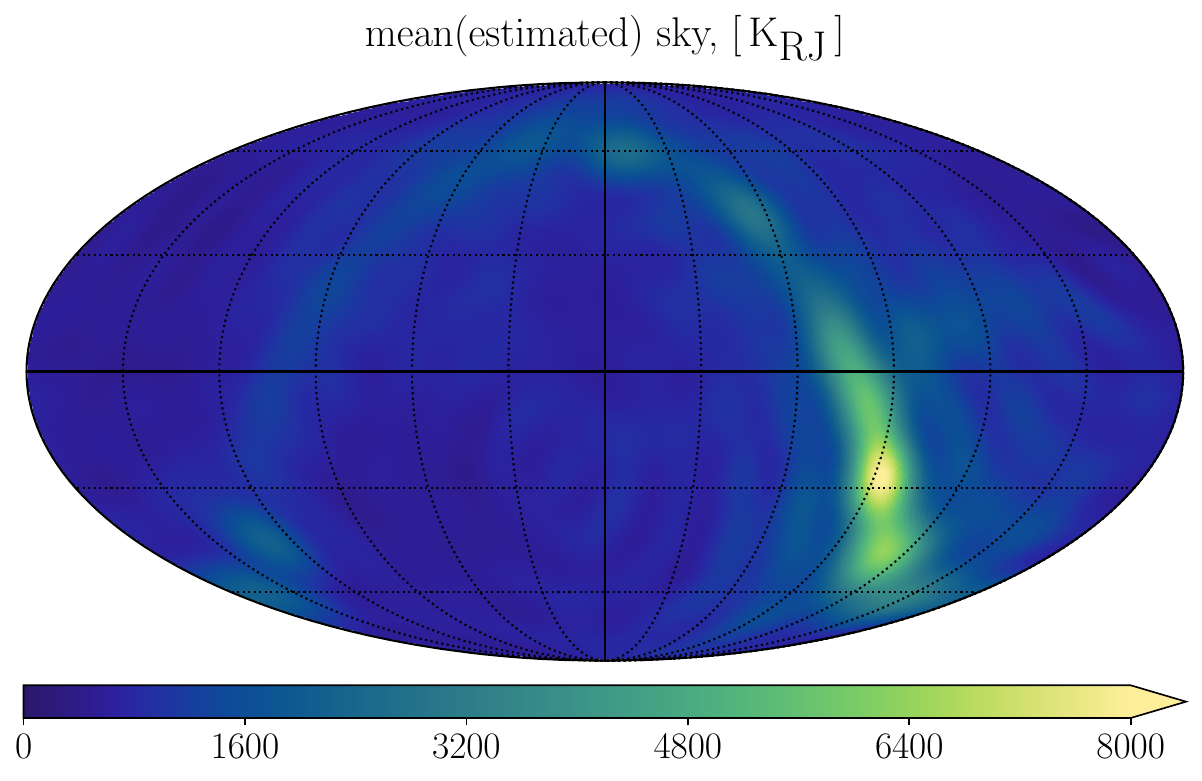}
    \includegraphics[width=0.95\columnwidth]{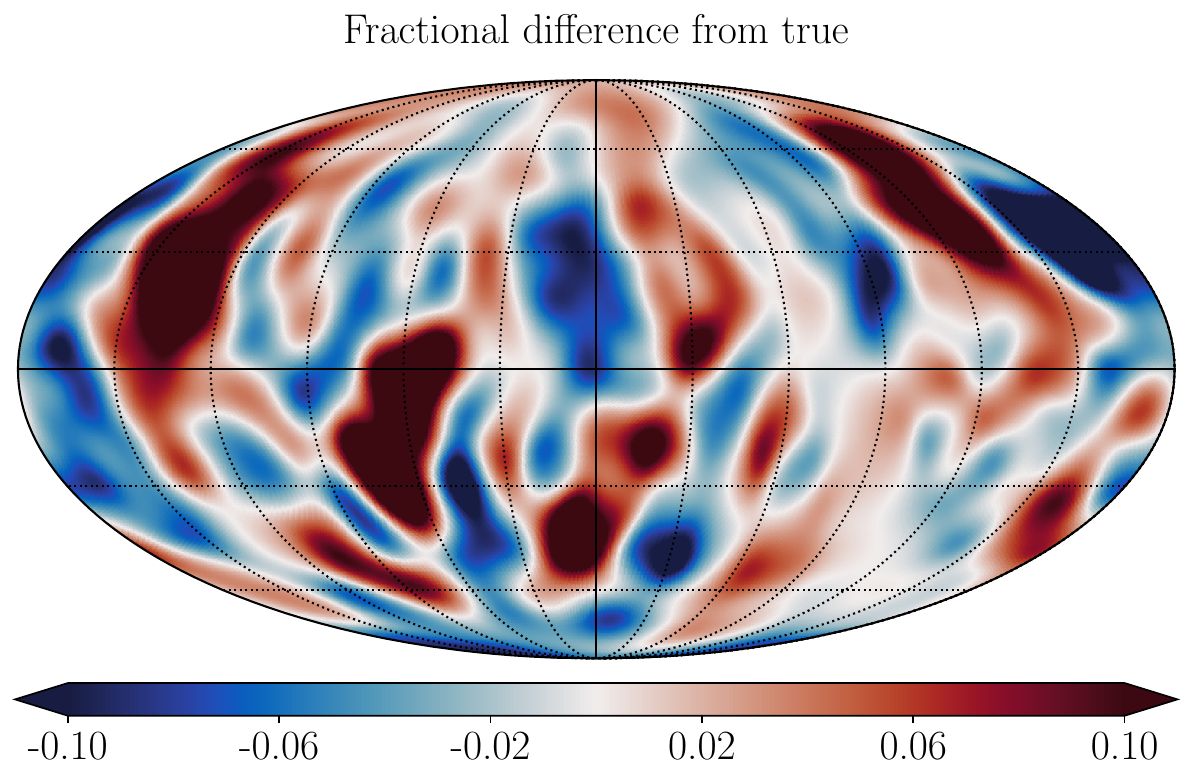}\hspace{0.06\columnwidth}
	\includegraphics[width=0.95\columnwidth]{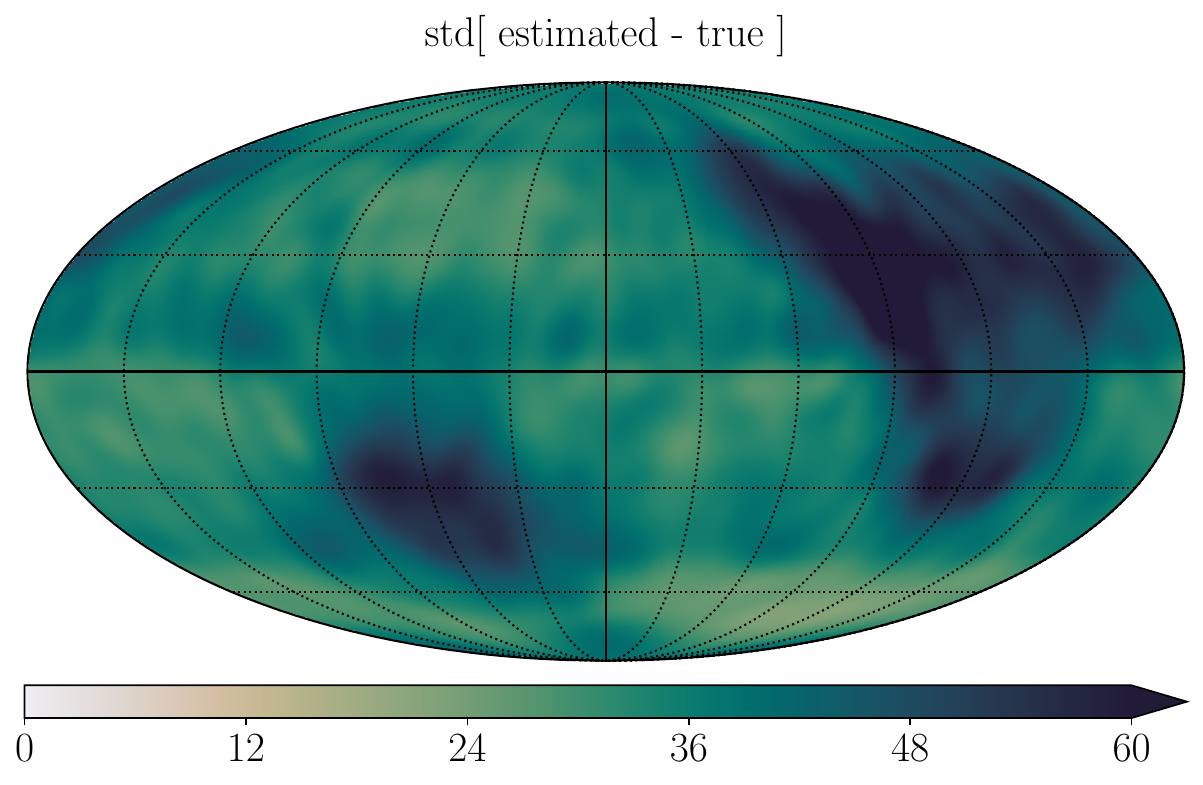}
    \caption{Resulting maps from the estimated spherical harmonic modes on the sky with the diameter of the receivers set to $\theta_{D}=\SI{3}{\meter}$ and thereby increasing the field of view. \emph{Upper left:} The true sky given by \texttt{pyGSM} with $\lmax=20$ and \nside$=128$. Note that the rippled structure comes from the true sky and not from the GCR solver. \emph{Upper right:} The estimated sky based on the mean of \comment{200} samples from the GCR solver. The spherical harmonic modes can be seen in Fig.~\ref{fig:alms_mwa}. \emph{Bottom left:} Fractional difference between the mean and the true sky  adjusted to show differences $<10\%$ (i.e. highlighting improvements over the prior). The beam region is no longer clear cut as in the standard case. \emph{Bottom right:} The standard deviation of the difference of the estimated and true sky. The lowest noise levels are now larger than with the standard case, but overall noise is lower.}
    \label{fig:sky_maps_mwa}
\end{figure*}

One reason that the smaller scales are more difficult to constrain comes from the array layout itself, as discussed above.
The prior helps breaking the degeneracy of the low-$\ell$ modes, that can otherwise only be determined as a linear combination (i.e. they are degenerate, as there are no sufficiently short baselines to resolve them individually). It is clear, however, from the \alm modes presented in Fig.~\ref{fig:alms_standard} that these modes are still difficult to constrain up to $\ell\sim 5$. Moving on the same logic as above, where the observable sky is increased to increase sensitivity to the larger scales, we can decrease the size of the dishes to increase the field of view (FoV). With the HERA-like dishes the FWHM $=\SI{12.3}{\degree}$. Altering the array to have a FoV more similar to that of OVRO-LWA \citep{Eastwood_2018} or MWA \citep{Yoshiura_2021}, the diameter of the receivers is changed to $\SI{3}{\meter}$, now with a ${\rm FWHM}=\SI{57.3}{\degree}$ (case~\case{D1}). A broader FoV now means that the smearing around the $\ell$ modes in Fig.~\ref{fig:baseline_vs_ell} is much narrower ($\delta\ell\simeq\pi/{\rm FWHM}$), thus more clearly picking out the specific $\ell$ modes.
Unlike when we increase the LST range, we expand the observable sky not only in the E-W direction but also N-S. 
As increasing the FoV with more than $4\times{\rm FWHM}$ of the standard case is a much more radical change than those made in Sec.~\ref{subsec:array_and_observing_configs}, the \case{D1}  results are presented in more detail in Figs.~\ref{fig:alms_mwa} and \ref{fig:sky_maps_mwa}, although now leaving out the Cartesian projections of the fractional residual, since the beam is now so wide that a close up of the primary beam region is no longer relevant. 
Starting with the true-subtracted mean of the recovered $\alm$ modes in Fig.~\ref{fig:alms_mwa}, it is noticeable that there are no longer any \emph{outliers} from the displayed region, as was the case in Fig.~\ref{fig:alms_standard} with the standard configuration. The zeroth-mode is still the most difficult to constrain and is again underestimated, however, not as severely underestimated as earlier. Earlier it was argued that the $\alm$ modes are not the most representative performance indicator of the code, since they are all-sky quantities and the observed region is only a narrow strip, but now that we have much larger coverage it clearly shows that full-sky features as the $\alm$ modes can be well-constrained. 

\comment{By visual inspection} it is now almost impossible to tell the difference between the true sky map and the mean of the recovered sky in Fig.~\ref{fig:sky_maps_mwa}. Taking the fractional difference between the two maps reveals a very different structure than earlier. Now, almost all parts of the sky are recovered within $10\%$ with the Galaxy showing the smallest residuals. It is striking that the primary beam structure no longer shows up in either the fractional difference map or the standard deviation. Since the noise covariance $\mathbf{N}$ and subsequently the noise $n_{ij}$ on the data is defined by the auto-visibilities as per Eq.~\ref{eq:radiometer_equation}, the noise level will be affected by whether the Galactic emission is within the beam. Since the FoV is now so large, the noise has increased with the extra power from the Galaxy as most of this falls within the observed area.

Additional runs have been made where the noise was reduced with a factor of $10^{-1}$ and $10^{-2}$ respectively (not shown), and not only does this improve the estimate of the $\alm$ modes, it is also clear that the area of the sky with the largest fractional residual is the furthest region from the observed region, corresponding to the same area in Fig.~\ref{fig:sky_maps_mwa} where the standard deviation is higher. Decreasing the noise also reduced the standard deviation inside the observed region to the same level as the standard case. Earlier, in Sec.~\ref{subsec:noise_and_prior_levels}, we showed that reducing the noise level could lead to worse recovery of the sky. Since we are dealing with constrained realizations, the solution in the prior-dominated region can still depend on the data-dominated region (and vice-versa), as the spherical harmonic basis functions connect the two. The influence of the data on the prior-dominated region is minimal if the noise covariance is very narrow. For the $\SI{3}{\meter}$ beam case it is not as big an issue, since the large FoV makes up for it. Before the beam was also very narrow, so when the noise covariance was made narrow, it did not allow the prior to contribute as much, which made it difficult to recover the sky even at the edges of the observed region. Now, that the FoV is larger, the prior is not as crucial to determine full-sky features and we can reduce the noise without losing the constraining power. 

\section{Conclusions}
\label{sec:conclusions}
For next-generation wide-field radio interferometer arrays, particularly those targeting the 21 cm fluctuations at low and high redshift, the biggest challenge remains proper handling of foreground contamination following its distortion/modulation by the instrument. To improve on this issue, it is crucial to be able to make accurate maps of the radio emission on both small and large scales from the measured visibilities. Traditional deconvolution algorithms such as CLEAN (and its extensions) can struggle to properly recover diffuse emission, which is the dominant foreground contaminant on large angular scales. To address this, methods such as Direct Optimal Mapping and $m$-mode analysis have been developed, both of which are maximum likelihood estimators that focus on accurately recovering wide field maps of the emission, i.e. on scales where the curvature of the sky and the shape of the primary beam becomes important.

In this work we have presented an alternative approach to recovering diffuse foregrounds from visibilities -- by constructing a parametric model of the emission, represented by spherical harmonics on the sky, and drawing samples from the joint posterior of the spherical harmonic coefficients given the visibility data and a choice of prior. A linear model can be constructed by writing down an operator that encodes the response of the interferometer (and thus the measured visibilities) to each spherical harmonic mode. This operator includes all the relevant instrumental degrees of freedom. We can then estimate the joint posterior distribution of all the coefficients by repeatedly solving the GCR equation, an extension of the Wiener filter that includes random realization terms. This can be solved efficiently even for a very large number of coefficients, making this large inference problem computationally tractable. Furthermore, each sample of coefficients is a complete realization of the spherical harmonic coefficients, and therefore the full sky (i.e. with no gaps or mask) that is statistically-consistent with the data. This ultimately allows us to recover the diffuse emission signal in a statistically robust manner while also avoiding difficulties with missing data in subsequent steps of the data analysis pipeline.

After presenting the mathematical formalism for this sampler, the primary aim of this paper was to validate the method by applying it to a suite of simulations. We tested the performance of the spherical harmonic sampler by comparing the recovered sky, defined as the mean of the sky maps of the samples, to the (known) true sky modelled using \texttt{pyGSM}. 
For the analysis we chose a standard set of parameters to act as a reference for the various tests of noise and prior uncertainty levels, sample size, $\lmax$, the specific LST range of the observations, and the influence of the field of view. The \emph{standard case} is based on \comment{200} realizations with a $10\%$ prior centred on the true $\alm$-modes, and the noise is given by a Gaussian distribution with zero mean and covariance defined by the auto-visibilities through the radiometer equation. The standard case uses a HERA-like closed-packed redundant array of 10 antennas and covers the LST range $0-\SI{8}{\hour}$. The dishes are 14 m in diameter (FWHM $=\SI{12.3}{\degree}$), which means the directly observed sky is a narrow $\SI{24.6}{\degree}$ (dec) by $\SI{120}{\degree}$ (RA) strip covering only $\sim8.6\%$ of the full sky.

The standard case performs well within the observed region and recovers the true sky to within $10\%$ within a distance of 1 FWHM from the centre and to within $5\%$ at the centre of the beam. 
When increasing the number of samples to $N_\textup{samples}=5000$, we found that there were minimal changes to the result allowing us to keep the computational costs lower by continuing with \comment{200} samples per scenario only.
\comment{The high-sample size case (\case{A1}) allowed us to also investigate the posterior distributions, which showed that the high-$\ell$ modes are consistent with the input value to within the statistical error. However, the low-$\ell$ modes (scales larger than the FoV) show smaller errorbars than we naively expected, despite having significant scatter around the true (input) values. We believe this behaviour to be an artefact of how these unresolved modes interact in order to fit the 'apparent' monopole within the directly observed patch of sky (which is different from the true monopole over the whole sky).}

Next, the impact of the noise level on recovery was studied. When the noise variance is increased by a factor of $>10^2$, the noise will dominate over the sky signal and downgrade the recovery. At this noise level, the samples will only show diffuse sky structure if there is a prior as well to drive the solution. 
For the standard case we chose a $10\%$ prior, as the \texttt{pyGSM} model should describe the diffuse emission sky to roughly within this margin. If the prior covariance is increased, however, the prior quickly becomes too broad and uncertain and the only well-recovered part is the very centre of the observed region.
Including a prior helps break the degeneracy of the low-$\ell$ modes. It was also found that even if the prior mean is set to zero (a highly pessimistic assumption), the sky can still be recovered, although to a slightly worse degree than the standard case. This shows the benefit of including a prior; the boundary region between observed and unobserved sky is constrained by both the prior and the data, while regions of missing data are still filled with a statistically plausible realization. We conclude that, for the standard case, the prior is not overly-informative, and contributes about the same as the data to the diffuse emission recovery around the observed region. 

For close-packed arrays like HERA, the most abundant baselines are the shortest ones, which have lengths $<\SI{30}{m}$. This justifies choosing a relatively low $\lmax$ for our tests. 
When increasing the $\lmax$ from 20 to 30, the number of modes in the $\alm$-vector increases significantly from 441 to 961, more than doubling the number of parameters we are trying to constrain, and thus stretching the available signal-to-noise across more degrees of freedom. A more informative prior would help to balance this increase in the size of the parameter space.

Most of the diffuse emission on the sky originates from the Galaxy and is not uniformly distributed. 
This results in field-dependent effects on the recovery of the signal. For instance, the recovery was best for a $8-\SI{16}{\hour}$ region, whereas the region covering the brightest part of the Galaxy ($16-24\si{\hour}$) was recovered worst, due largely to the extra noise given by the auto-visibilities.
When increasing the data volume we expect the results to improve, for instance if we increase the cadence and take more data samples within the same LST range, or, if we keep the cadence but increase the range. When doubling the number of data points within the range, there was only a slight improvement to the recovered sky, suggesting that there is not much information to be gained by making more measurements of nearby (correlated) pointings, e.g. within a beam width of one another. In contrast, doubling the range itself not only improved the results in the centre, but also at the boundaries of the LST range, as additional sky coverage led to improved constraints on the spherical harmonic modes across the board.

Ultimately, the biggest impact on our simulated results occurred by increasing the field of view. This was done by decreasing the effective antenna diameter to $\SI{3}{\meter}$, thus increasing the FWHM from $\SI{12.3}{\degree}$ to $\SI{57.3}{\degree}$. 
For the large-FoV case, the spherical harmonic coefficients
now all coincide with the true sky within their given uncertainties except the zeroth mode. The standard deviation map of the large-FoV results no longer shows the clear strip of the observed sky, as almost all of the modes are now well-constrained. Since the noise is defined by the auto-visibilities, and this setup now directly observes most of the Galaxy, the noise level is larger than the standard case. However, as the FoV is now so large, we are also less dependent on the prior as there are fewer gaps of information to fill in, and the boundary region (between directly observed and totally unobserved regions) is wider. 

The results presented in this paper are based on a handful of simplified example cases, using a reduced-size array and only two frequencies for the visibility response operator. A simple extension of this work would be to rerun the analysis while simulating the visibility response for more frequencies or, alternatively, the frequency dependence could be included via a parametric model, e.g. power law with spectral index $\beta$. Likewise, we have only considered cases of $\lmax\comment{\leq}30$, but in reality a higher angular resolution would be desirable if a larger portion of the array is considered (i.e. including more longer baselines). 
\comment{In real data effects like cable reflections, uncleaned RFI, polarisation leakage, cross-talk, etc. are also present. If these are not included in the data model there is a risk of biasing the inferred field as it will try to absorb the unmodelled features. The solution to this is either to use a more sophisticated data model or include a flexible systematics term to account for these features.}

Finally, the GCR method has been developed with inclusion into a Gibbs sampling framework in mind, whereby beam, 21cm signal, and point source foreground model parameters would also be sampled. A simpler and more focused Gibbs sampling scheme would also enable us to sample the signal covariance $\vecb{S}$ (e.g. as parametrised by the angular power spectrum), as well as the spherical harmonic coefficients themselves.

\section*{Acknowledgements}

This result is part of a project that has received funding from the European Research Council (ERC) under the European Union's Horizon 2020 research and innovation programme (Grant agreement No. 948764; KAG, PB, JB, MJW). 
We acknowledge use of the following software: 
{\tt matplotlib} \citep{matplotlib}, {\tt numpy} \citep{numpy}, and {\tt scipy} \citep{2020SciPy-NMeth}.  This work used the DiRAC@Durham facility managed by the Institute for Computational Cosmology on behalf of the STFC DiRAC HPC Facility (www.dirac.ac.uk). The equipment was funded by BEIS capital funding via STFC capital grants ST/P002293/1, ST/R002371/1 and ST/S002502/1, Durham University and STFC operations grant ST/R000832/1. DiRAC is part of the National e-Infrastructure.

\section*{Data Availability}

The Python code used to produce the results in this paper is implemented in \url{https://github.com/HydraRadio/Hydra} and is also available from \url{https://github.com/katrinealice/sph_harm_GCR}. The Jupyter notebook used for the data analysis and to produce plots is available from \url{https://github.com/katrinealice/sph_harm_GCR_analysis}. Simulated data are available on request.
 



\bibliographystyle{mnras}
\bibliography{bibliography} 




\appendix

\balance

\section{Further ``realification''}
\label{app:realification}

Inherently visibilities are complex quantities and the spherical harmonic coefficients are as well. Many standard linear solvers are not set up to handle complex numbers however. The first step to make the system real (i.e. \emph{realify}) has already been done in Sec.~\ref{subsec:data_model} bt reordering the spherical harmonic coefficient vector, $\vecb{a}_{\ell m}$, to contain the values of first the real- and then the imaginary part of the spherical harmonic coefficients. This leaves us with a fully real-valued vector to solve for, denoted $\vecb{a}_{\rm cr}$ (constrained realizations). A real-valued spherical harmonic coefficient vector does not automatically result in real-valued visibilities however. Instead, the visibility response is still complex valued, now containing a complex visibility response per real- and per imaginary part of the $\vecb{a}_{\rm cr}$-vector. Further \emph{realification} of the system is therefore needed, while making sure that all the mixing of real and imaginary parts that you get when multiplying complex numbers together is handled correctly.

First, we define vectors (bold, lower case) and matrices (bold, upper case) in this real-valued system as
\begin{align}
\widetilde{\vecb{v}}=\begin{pmatrix}
    \vecb{v}_{\rm re} \\ \vecb{v}_{\rm im}
\end{pmatrix}, \quad
\widetilde{\vecb{M}} = \begin{pmatrix}
    \vecb{M}_{\rm re} & -\vecb{M}_{\rm im} \\[3pt] 
    \vecb{M}_{\rm im} & \vecb{M}_{\rm re}
\end{pmatrix}
\end{align}
with the Hermitian conjugate of the matrix $\widetilde{\vecb{M}}$ given as
\begin{align}
\widetilde{\vecb{M}}^{\dag} = \begin{pmatrix}
    \vecb{M}_{\rm re}^T & \vecb{M}_{\rm im}^T \\[3pt]
    -\vecb{M}_{\rm im}^T & \vecb{M}_{\rm re}^T
\end{pmatrix},
\end{align}
where the superscript $T$ denotes the transpose and the minus sign has been swapped due to complex conjugation. 

With this we can redefine the complex-valued visibility response $\vecb{X}$ into a purely real-valued version,
\begin{align}
\widetilde{\vecb{X}} = \begin{pmatrix}
    \vecb{X}_{\textup{re}} & -\vecb{X}_{\textup{im}} \\[3pt] 
    \vecb{X}_{\textup{im}} & \vecb{X}_{\textup{re}}
\end{pmatrix},\quad
\widetilde{\vecb{X}}^\dag = \begin{pmatrix}
    \vecb{X}_{\textup{re}}^T & \vecb{X}_{\textup{im}}^T \\[3pt] 
    -\vecb{X}_{\textup{im}}^T & \vecb{X}_{\textup{re}}^T
\end{pmatrix}\,,
\end{align}
and we define the new noise covariance as a diagonal matrix
\begin{align}
\widetilde{\vecb{N}} = \begin{pmatrix}
    \vecb{N}/2 & 0 \\ 0 & \vecb{N}/2
\end{pmatrix}\,.
\end{align}

Since the spherical harmonic coefficient vector is already constructed so that it is real-valued, as described in Sec.~\ref{subsec:data_model}, so is the prior mean $\vecb{a}_0$ and $\vecb{\omega}_a$,
\begin{align}
\widetilde{\vecb{a}}=\begin{pmatrix}
    \vecb{a}_{\rm cr} \\ 0
\end{pmatrix},\quad
\widetilde{\vecb{a}}_0=\begin{pmatrix}
    \vecb{a}_{0} \\ 0
\end{pmatrix},\quad
\widetilde{\vecb{\omega}}_a=\begin{pmatrix}
    \vecb{\omega}_{a} \\ 0
\end{pmatrix}\,.
\end{align}
Lastly, the signal covariance is defined in Sec.~\ref{subsec:data_and_noise_model} by the real-valued $\vecb{a}_\text{true}$-vector and so, already, is real-valued itself. We therefore define the new signal covariance as,
\begin{align}
    \widetilde{\vecb{S}} = \begin{pmatrix}
    \comment{ \vecb{S}} & 0 \\ 0 & 0
\end{pmatrix}\,.
\end{align}

Using these definitions, we can derive the GCR equation once again, and we are left with the final \emph{realified} GCR equation,
\begin{align}
    \left[\vecb{S}^{-1} + 2\vecb{X}_{\textup{re}}^T \vecb{N}^{-1} \vecb{X}_{\textup{re}} + 2\vecb{X}_{\textup{im}}^T\right.&\left. \vecb{N}^{-1} \vecb{X}_{\textup{im}}\right] \vecb{a}_{\textup{cr}} =\nonumber\\ 
    &\quad \vecb{X}_{\textup{re}}^T \left(2\vecb{N}^{-1}\vecb{d}_{\textup{re}} + \sqrt{2}\vecb{N}^{-\frac{1}{2}} (\vecb{\omega}_d)_{\textup{re}}\right)\nonumber\\ 
    &\quad + \vecb{X}_{\textup{im}}^{\comment{T}} \left( 2\vecb{N}^{-1} \vecb{d}_{\textup{im}} + \sqrt{2}\vecb{N}^{-\frac{1}{2}} (\vecb{\omega}_d)_{\textup{im}} \right) \nonumber\\
    &\quad + \vecb{S}^{-1} \vecb{a}_0 + \vecb{S}^{-\frac{1}{2}} \vecb{\omega}_a .
    \label{eq:realified_GCR_real}
\end{align}



\bsp	
\label{lastpage}
\end{document}